\documentclass{elsart}

\usepackage{natbib}

\usepackage{graphicx}

\usepackage{amsmath, amsfonts, amssymb}

\newcommand{\ie}{\textit{i.e.}~}
\newcommand{\eg}{\textit{e.g.}~}
\newcommand{\etc}{\textit{etc.}}
\newcommand{\MSun}{M_{\odot}}
\newcommand{\MEarth}{M_{\oplus}}

\begin{document}

\begin{frontmatter}



\title{Planetesimal-driven Planet Migration in the Presence of a Gas Disk}


\author[queens]{Christopher C. Capobianco}, 
\author[queens]{Martin Duncan}, and
\author[swri]{Harold F. Levison}

\address[queens]{Department of Physics, Engineering Physics and Astronomy, Queen’s University, 99 University Avenue, Kingston, Ontario, K7L 3N6, Canada}
\address[swri]{Southwest Research Institute, Department of Space Studies, 1050 Walnut Street Suite 429, Boulder, CO 80302, USA}

\begin{center}
\scriptsize
Copyright \copyright 2010 Christopher C. Capobianco, Martin Duncan and Harold F. Levison\\
Accepted for publication in Icarus: September 1, 2010
\end{center}


\begin{abstract}

We report here on an extension of a previous study \citet{2009Icar..199..197K} of planetesimal-driven migration using our $ N $-body code SyMBA \citep{1998AJ....116.2067D}.  The previous work focused on the case of a single planet of mass $ M_{\mathrm{em}} $, immersed in a planetesimal disk with a power-law surface density distribution and Rayleigh distributed eccentricities and inclinations.  Typically $ 10^4 $ to $ 10^5 $ equal-mass planetesimals were used, where the gravitational force (and the back-reaction) on each planetesimal by the Sun and planet were included, while planetesimal-planetesimal interactions were neglected.  The runs reported on here incorporate the dynamical effects of a gas disk, where the \citet{1976PThPh..56.1756A} prescription of aerodynamic gas drag is implemented for all bodies.  In some cases the \citet{2000MNRAS.315..823P} prescription of Type -- I migration for the planet are implemented, as well as a mass distribution.

In the gas-free cases, rapid planet migration was observed -- at a rate independent of the planet's mass -- provided the planet's mass was not large compared to the mass in planetesimals capable of entering its Hill sphere.  In such cases, both inward and outward migrations can be self-sustaining, but there is a strong propensity for inward migration.  When a gas disk is present, aerodynamic drag can substantially modify the dynamics of scattered planetesimals. For sufficiently large or small mono-dispersed planetesimals, the planet typically migrates inward. However, for a range of plausible planetesimal sizes (\ie 0.5 -- 5.0 km at 5.0 AU in a minimum mass Hayashi disk) outward migration is usually triggered, often accompanied by substantial planetary mass accretion.  The origins of this behaviour are explained in terms of a toy model.  The effects of including a size distribution and torques associated with Type -- I migration are also discussed.

\end{abstract}



\end{frontmatter}

\section{Introduction\label{intro}}

Currently, over 460 extrasolar planets are known\footnote{See http://exoplanet.eu}, along with over 40 systems containing multiple planets.  Most of the extrasolar planets detected to date have masses comparable to that of Neptune, or larger.  Furthermore in a recent summary by \citet{2007prpl.conf..685U}, at least $ \sim 6\% $ of stars surveyed have giant planets interior to $ \sim 5.0 \, \hbox{AU} $, so giant planets appear fairly common in stellar systems.  Moreover, results from the HARPS survey \citet{2008A&A...487..373S} shows that Neptune-mass extrasolar planets are found in $ \sim 40\% $ of the stars surveyed.  Most likely, these giant planets formed via a similar process that formed the four giant planets in the Solar System.

However, the masses and orbits of these extrasolar planets display a wide variety of configurations: \eg Neptune and Jupiter-mass planets with short orbital periods, isolated planets with large orbital eccentricities, multiple planet systems in resonance, and planets orbiting components of stellar binaries.  Several analytical models have been proposed to explain the various aspects of planet formation, but most of these have not been tested numerically.  Until recently, very little had been done on giant planet core formation using $ N $-body simulations \citep{2003Icar..161..431T}.  \citet{2010AJ....139.1297L} (hereafter referred to as LTD10) recently completed a comprehensive set of computer simulations which included a number of physical processes that might enhance accretion onto planetary embryos. As discussed in Section 2, the most successful models were those in which one or more embryos spontaneously underwent a burst of outward migration induced by planetesimal scattering.

In an attempt to further our understanding of some of the results in LTD10, we are undertaking a detailed investigation of the combined effects of planetesimal scattering and aerodynamic drag on the growth and evolution of giant planet cores.  Our goal in this paper is to understand the case of the dynamics of a single core interacting with a disk of planetesimals and gas.  In what follows, we provide some background in \S\ref{planet_form}, then briefly discuss our implementation of the relevant forces in \S\ref{circum_processes}.  In \S\ref{simulations} we discuss the results of simulations including aerodynamic gas drag, for a disk of mono-dispersed planetesimals.  A toy model which explains the results is presented in \S\ref{toy_model}.  The effects of a planetesimal size distribution is presented in \S\ref{size_distribution}, and Type -- I migration in \S\ref{type_I_effects}.  A summary and conclusion is presented in \S\ref{conclusions}.

\section{Giant Planet Formation\label{planet_form}}

The formation of giant planets in the widely adopted core accretion model can typically be described in four stages.  The first stage involves the formation of planetesimals, which we do not model in this study.  The next stage involves the runaway accretion of planetesimals by a small fraction of those planetesimals which happen to grow a bit larger, and then grow much faster than all the others \citep{1989Icar...77..330W}.  When these large bodies become sufficiently massive and well-spaced such that each dominates the viscous stirring in its feeding zone, the runaway growth gives way to the oligarchic growth stage.  An embryo's feeding zone is the annulus about its orbit where small bodies can suffer strong gravitational impulses.  Typically this feeding zone extends from 1.0 -- 3.5 Hill radii on either side of the embryo's orbit, and is the source of most of the material which the embryo accretes.  During the oligarchic stage, the large embryos grow in lockstep, maintaining similar masses and uniformly spaced orbits \citep{1998Icar..131..171K, 2003Icar..161..431T}.  The final stage in the outer solar system is characterized by the rapid accumulation of a gaseous envelope by the embryos; in the inner region it is characterized by the giant impact phase of terrestrial planet formation.

However, the core accretion model has its weaknesses.  In particular, the accretion of a massive atmosphere requires a solid core of mass $ \sim 10 \, \MEarth $ to trigger a rapid gas accretion phase \citep{1980PThPh..64..544M, 1996Icar..124...62P, 2005Icar..179..415H}. The difficulties of reaching this threshold are threefold:
\begin{enumerate}
	\item Accretion has to be sufficiently efficient to concentrate enough mass into at least one body, and potentially multiple bodies.
	\item Accretion has to occur within $ \sim 10 \, \hbox{Myr} $ \citep{2001ApJ...553L.153H}, such that there is $ \sim 10^2 \, \MEarth $ left in the nearby disk to furnish an envelope.
	\item Migration due to embryo-disk tidal interactions (cf. \S\ref{type_I_migration}), threatens to deposit core-sized bodies into the central star faster than they can accrete \citep{1986Icar...67..164W, 1993Icar..102..150K, 1997Icar..126..261W}.
\end{enumerate}

Several analytical models have been proposed to mitigate these problems, and some of these have been tested numerically by LTD10.  In particular, LTD10 numerically integrated the orbits of a number of planetary embryos embedded in a swarm of planetesimals. Their simulations included simplified models of various combinations of the following effects:  (1) aerodynamic drag on small bodies, (2) collisional damping, (3) extended atmospheres around the embryos \citep{2003A&A...410..711I}, (4) embryo eccentricity damping due to gravitational interaction with the gas disk, (5) fragmentation of the planetesimals and (6) evaporation and re-condensation at the snow line \citep{2004ApJ...614..490C}.  They found that the gravitational interaction between the embryos and the planetesimals generally led to regions near the embryos being cleared of planetesimals before much accretion onto the embryos could occur. However, the most successful phases of embryo growth occurred when the gravitational scattering of the planetesimals, together with the effects of aerodynamic gas drag led to the rapid outward migration of one or more embryos.  We show in this paper that many of the main features of the embryo-planetesimal interactions that lead to rapid outward migration and planet growth are demonstrated by the single embryo case which we discuss next.

\section{Physical Processes in Circumstellar Disks\label{circum_processes}}

There are several physical processes that can occur in circumstellar disks; some of these are only relevant to planetesimals, while others only to larger embryo-sized bodies.  Specifically, the dynamics of planetesimals and embryos will be affected by the gravitational perturbations from other massive bodies, as well as gas effects.  Radiative forces are not very important for 0.01 -- 100 km size bodies over the timescale under consideration (\ie $ \sim 10 \, \hbox{Myr} $), so we neglect such forces in the subsequent discussion.

\subsection{Gravitational Effects\label{grav_effects}}

The dominant gravitational influence in the circumstellar environment, for planetesimals and embryos, is the central star.  However, in the vicinity of other massive bodies (\eg embryos), the gravitational tidal influence of those massive bodies will dominate.  The transition is characterized by a length scale called the Hill radius, which defines a sphere about each body where its gravitational tide dominates the gravitational influence from the central star:

\begin{equation}
	R_h \equiv a \left( \dfrac{M}{3M_{\star}} \right )^{1/3}
	\label{eq:hill_radius}
\end{equation}
\noindent where $ M $ and $ a $ are the mass and the semi-major axis of an orbiting body, while $ M_{\star} $ is the mass of the central star.  At 1.0 AU an Earth-mass object would have $ R_h \simeq 0.01 \, \hbox{AU} $, while at 10.0 AU a Jupiter-mass object would have $ R_h \simeq 0.7 \, \hbox{AU} $.

In the event of a close encounter, the embryo will tend to scatter the planetesimal to a smaller or larger orbit, exchanging energy and angular momentum.  Consequently, the embryo will respond by moving in the opposite direction of the planetesimal, albeit by a much smaller amount.  Since an embryo is surrounded by a swarm of planetesimals, it will scatter numerous planetesimals as it moves along its orbit.  Furthermore, if the probability of scattering a planetesimal inwards were the same as scattering outwards, there will be no net change of the embryo's orbit.  However, since the timescale for a scattering encounter is slightly shorter inside the planet's orbit, it will preferentially scatter planetesimals from inside its orbit to outside its orbit.  Consequently, the embryo will experience a net inward drift, and this inward migration will continue so long as there is sufficient material for it to scatter \citep{1984Icar...58..109F, 1993Natur.365..819M, 2004Icar..170..492G}.  This migration is studied in detail by \citet{2007PhDT........21K} and \citet{2009Icar..199..197K} in gas-free disks, and we briefly summarize their work here.

In their study \citet{2009Icar..199..197K} noted that if a swarm of planetesimals were scattered by a much more massive embryo, it could lead to a net exchange of angular momentum that would induce the embryo to migrate.  The rate an embryo's orbital distance will drift due to planetesimal scattering is given by \citep{2009Icar..199..197K}:

\begin{equation}
	\left . \frac{\dot{a}}{a}\right |_{\mathrm{sca}} \simeq -\dfrac{2}{P_{\mathrm{orb}}} \left ( \dfrac{M_{\mathrm{disk}}}{\MSun} \right ) \left [ 1 + \dfrac{1}{5} \left ( \dfrac{M_{\mathrm{em}}}{M_{\mathrm{enc}}} \right )^3 \right ]^{-1}
	\label{eq:adot_scatter}
\end{equation}
\noindent where $ P_{\mathrm{orb}} $ is the embryo's orbital period at $ a_{\mathrm{em}} $, while $ M_{\mathrm{disk}} \equiv \Sigma_{\mathrm{solid}}(a_{\mathrm{em}}) \pi a_{\mathrm{em}}^2 $ is the local mass of the disk, where $ \Sigma_{\mathrm{solid}}(a_{\mathrm{em}}) $ is the local surface density of the solid material in the disk and $ \MSun $ is the solar mass.  This rate will be independent of $ M_{\mathrm{em}} $ provided $ M_{\mathrm{em}} \ll M_{\mathrm{enc}} $ where $ M_{\mathrm{enc}} \simeq 5 \xi_h M_{\mathrm{disk}} $ is the mass in the embryo's encounter region, and $ \xi_h = R_h/a_{\mathrm{em}} \equiv (M_{\mathrm{em}}/3M_{\star})^{1/3} = 10^{-2}(M_{\mathrm{em}}/\MEarth)^{1/3} $ is the Hill factor.  This rate is valid provided that the planetesimal eccentricities $ e \lesssim 3 \xi_h $, which is always the case considered in this paper.

\subsection{Gas Effects\label{gas_effects}}

\subsubsection{Aerodynamic Gas Drag\label{aero_gas_drag}}

While the gaseous component of the disk is still present, solid particles will be affected by an aerodynamic gas drag.  The strength of this gas drag acceleration is inversely proportional to a particle's radius, so larger particles will be less effected.  Since embryos can have radii greater than a few hundred km (depending on their density), gas drag is not an important effect.

The dynamics induced by aerodynamic gas drag on a particle is described by:

\begin{equation}
	\dot{\mathbf{v}}_{\mathrm{aero}} = \dfrac{\mathbf{v} - \mathbf{v}_{\mathrm{gas}}}{\tau_{\mathrm{aero}}}
	\label{eq:vdot_aero}
\end{equation}
\noindent where $ \mathbf{v} $ and $ \mathbf{v}_{\mathrm{gas}} $ are the velocity vectors of the planetesimal and the gas, and $ \tau_{\mathrm{aero}} $ is the aerodynamic gas drag timescale given by \citet{1976PThPh..56.1756A}:

\begin{equation}
	\tau_{\mathrm{aero}} = \dfrac{8\rho_{\mathrm{plsml}}R_{\mathrm{drag}}}{3C_D\rho_{\mathrm{gas}}v_{\mathrm{kep}}}
	\label{eq:tau_aero}
\end{equation}
\noindent where $ \rho_{\mathrm{plsml}} $ and $ R_{\mathrm{drag}} $ are the mass density and radius of the planetesimal, while $ \rho_{\mathrm{gas}} $ is the volume gas density and $ v_{\mathrm{kep}} $ is the Keplerian orbital velocity.  The drag coefficient $ C_D $ is a quantity often of order unity, and in general is a nonlinear function of the particle's radius and its relative velocity to the gas.  We discuss the behaviour of $ C_D $ in more detail below.

For the analytical discussion in the subsequent sections, it is useful to define damping timescales for planetesimals with small eccentricity $ e $ and inclination $ I $.  Damping rates for $ e $ and $ I $ of the planetesimals due to aerodynamic gas drag were calculated by \citet{1976PThPh..56.1756A}:

\begin{subequations}
	\label{eq:eidot_aero_drag}
	\begin{align}
		\left . \frac{\dot{e}}{e}\right |_{\mathrm{aero}} &= -\frac{1}{\tau_{\mathrm{aero}}}\sqrt{\eta^2+\frac{5}{8}e^2+\frac{1}{2}I^2} \label{eq:edot_aero_drag} \\
		\left . \frac{\dot{I}}{I}\right |_{\mathrm{aero}} &= -\frac{1}{2\tau_{\mathrm{aero}}}\sqrt{\eta^2+\frac{5}{8}e^2+\frac{1}{2}I^2} \label{eq:idot_aero_drag}
	\end{align}
\end{subequations}

\noindent where $ \eta \equiv \frac{1}{2} \left[ 1 - \left(\frac{v_{\mathrm{gas}}}{v_{\mathrm{kep}}}\right)^2\right]$ defines the deviation of the gas velocity from the Keplerian velocity at a given location in the disk.  Assuming a power-law volume density profile [\ie $ \rho_{\mathrm{gas}}(a) \propto a^{-\alpha} $], and a power-law gas temperature profile [\ie $ T_{\mathrm{gas}}(a) \propto a^{-1/2} $] in the disk gives: $ \eta(a) = 6.0 \times 10^{-4} \, (\alpha + \frac{1}{2}) (a/\hbox{AU})^{1/2} $.

The corresponding damping rate for a planetesimal's orbit is given by:

\begin{equation}
	\left . \frac{\dot{a}}{a}\right |_{\mathrm{aero}} = -\frac{2}{\tau_{\mathrm{aero}}}\sqrt{\eta^2 + \frac{5}{8}e^2 + \frac{1}{2}I^2} \left ( \eta + e^2 + \frac{1}{8}I^2 \right )
	\label{eq:adot_aero_drag}
\end{equation}

We now digress temporarily to discuss two quantities that appear in the gas drag formulae which are often overlooked.  The first is the quantity $ \eta^2 $, which as it appears in Eq.~\ref{eq:eidot_aero_drag} and Eq.~\ref{eq:adot_aero_drag}, is often significantly smaller compared to values of $ e^2 $ and $ I^2 $.  As a consequence, many studies that include these aerodynamic gas drag formulae often employ versions of these equations in the limit when $ \eta^2 \ll e^2, I^2 $.  However as the aerodynamic drag damps the random motion of the planetesimals particularly for small embryo masses and at larger distances in the disk, the value of $ \eta^2 $ will become comparable to $ e^2 $ and $ I^2 $.  It seems only prudent to implement the full equations from \citet{1976PThPh..56.1756A}, rather than their typical approximations.

\begin{figure}[ht!]
	\begin{center}
	\includegraphics[width = \columnwidth]{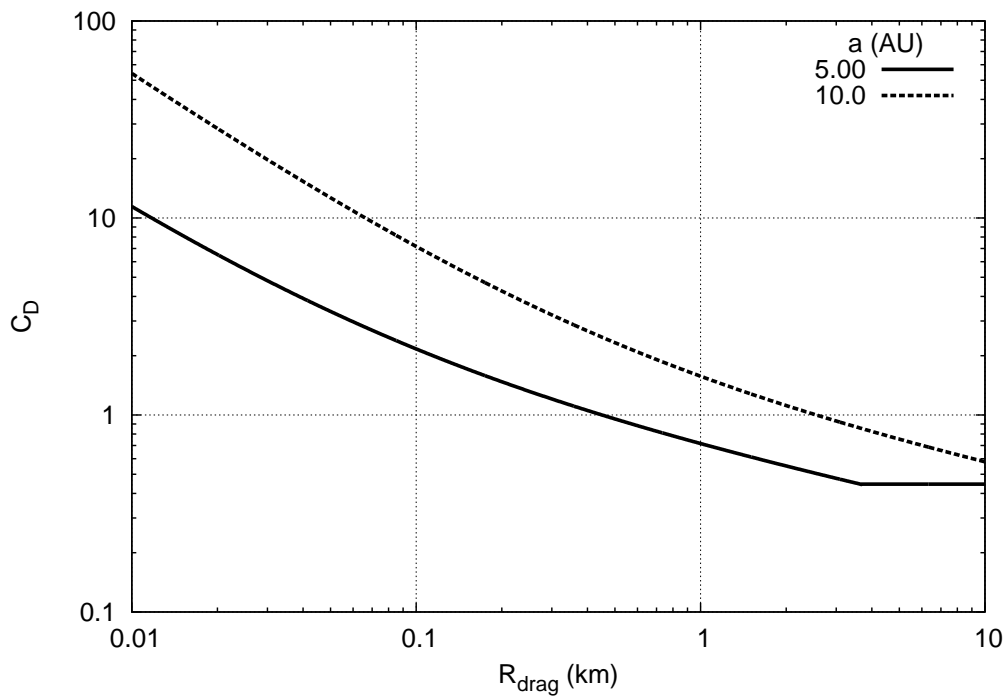}
	\caption[cd_rdrag_coefficient]{\label{fig:cd_rdrag_coefficient}
		Behaviour of the drag coefficient $ C_D $ as a function of planetesimal radius $ R_{\mathrm{drag}} $, assuming a gas volume density of $ 1.4 \times 10^{-9} \, \hbox{g cm}^{-3} $ at 1.0 AU which varies radially as a power-law: $ \rho_{\mathrm{gas}} \propto a^{-11/4} $.  It is also assumed that the planetesimal density $ \rho_{\mathrm{plsml}} = 0.5 \, \hbox{g cm}^{-3} $, which is the same in all the figures.  The two different curves correspond to two locations in the disk: 5.0 AU (solid line) and 10.0 AU (dashed line).}
	\end{center}
\end{figure}

Next is the quantity $ C_D $, which is often assumed to be of order unity, and is often fixed throughout the disk.  However, \citet{1976PThPh..56.1756A} and \citet{2004AJ....128.1348R} clearly showed that the value of $ C_D $ is a nonlinear function of the planetesimal radius and its velocity relative to the gas.  Furthermore, the value of $ C_D $ not only varies over a couple orders of magnitude across several orders of magnitude of planetesimal radius, but it also varies with distance in the disk.  This is illustrated in Fig.~\ref{fig:cd_rdrag_coefficient} which plots the value of $ C_D $ as a function of $ R_{\mathrm{drag}} $ at 5.0 AU and 10.0 AU, and assumes that a gas volume density varies radially as a power-law: $ \rho_{\mathrm{gas}} = 1.4 \times 10^{-9} \, \hbox{g cm}^{-3} \, (a/\hbox{AU})^{-11/4} $ as prescribed by the Minimum Mass Solar Nebula (MMSN) model \citep{1981PThPS..70...35H}.  The nonlinear treatment of $ C_D $ is incorporated in the work of \citet{2004AJ....128.1348R} and \citet{2007Icar..191..413B}, and for the purposes of our study we implement the formulae in \citet{2007Icar..191..413B}.

\subsubsection{Type -- I Migration\label{type_I_migration}}

For embryos, a more important dynamical effect than aerodynamic gas drag is the gravitational interaction with the gas disk itself.  Embryos embedded in a gas disk will launch density waves at their inner and outer Lindblad resonances (which are analogous to mean motion resonances in disks), and as a result experience a positive and negative torque from the inner and outer portions of the density wave, respectively \citep{1980ApJ...241..425G}.  Since the outer torque dominates in locally isothermal disks, this causes the embryo's orbit to decay, which is usually termed Type -- I migration \citep{1997Icar..126..261W}.  The migration rate increases with embryo mass until the embryo is large enough to clear a gap in the gas disk, at which point it becomes locked into the slower viscous evolution of the disk (\ie Type -- II migration, cf. \citet{1997Icar..126..261W}).  However, since embryos do not form a gap until they reach a mass of 10 -- $ 100 \, \MEarth $, then Type -- I migration will dominate embryo evolution until the embryo becomes a gas giant core.

Recent work has revealed that if the gas is not isothermal, then the drag from the gas in the horseshoe region can lead to outward driven Type -- I migration \citep{2008A&A...485..877P, 2010MNRAS.401.1950P}.  However for the radial density (\ie $ \rho_{\mathrm{gas}}(a) \propto a^{-\alpha}$) and temperature (\ie $ T_{\mathrm{gas}}(a) \propto a^{-1/2}$) profiles explored in this study, the simple torque formula \citep[cf.][Eq.~47]{2010MNRAS.401.1950P} indicates that the embryo's preferred migration direction would still be inward.  Work by \citet{2010ApJ...715L..68L} extends the analysis by including a viscously and radiatively evolving disk, which is an important improvement as most disk models are assumed static or isothermal.  While their simulations do indeed show regimes of outward migration, these regimes are typically circumscribed between equilibrium radii (\eg locations in the disk where the torque is zero) which drift inwards as the disk evolves.  They find that smaller mass embryos are able to decouple from the evolution of these equilibrium radii, but since they do not incorporate mass accretion for the embryo this may have limited application to our study.  

For Type -- I migration, \citet{2000MNRAS.315..823P} give the acceleration on an embryo due to the tidal effects

\begin{equation}
	\dot{\mathbf{v}}_{\mathrm{tidal}} = -\dfrac{\mathbf{v}}{\tau_{\mathrm{a,typeI}}} - 2\dfrac{\mathbf{v}\cdot\mathbf{r}}{r^2 \tau_{\mathrm{e,typeI}}} - 2\dfrac{\mathbf{v}\cdot\mathbf{k}}{\tau_{\mathrm{I,typeI}}}\mathbf{k}
	\label{eq:vdot_type_I}
\end{equation}
\noindent where $ \mathbf{r} $ and $ \mathbf{v} $ are the position and velocity vectors of the embryo, and $ \mathbf{k} $ is the unit vector in the vertical direction, while $ \tau_{\mathrm{a,typeI}} $, $ \tau_{\mathrm{e,typeI}} $ and $ \tau_{\mathrm{I,typeI}} $ are respectively the decay timescales for the orbital distance, eccentricity and inclination:

\begin{equation}
	\tau_{\mathrm{a,typeI}} = \dfrac{P_{\mathrm{orb}}}{2 \pi c_a} \, \left ( \dfrac{z_s}{a} \right )^{2} \left ( \dfrac{\Sigma_{\mathrm{gas}} \pi a^2}{\MSun} \right )^{-1} \, \left ( \frac{M_{\mathrm{em}}}{\MSun} \right )^{-1} \, \left [ \dfrac{1 + ( \frac{e a}{1.3 z_s } )^{5}}{1 - ( \frac{e a}{1.1 z_s } )^{4}} \right ]
	\label{eq:tau_a_type_I}
\end{equation}

\begin{equation}
	\tau_{\mathrm{e,typeI}} = \dfrac{P_{\mathrm{orb}}}{2 \pi c_e} \, \left ( \dfrac{z_s}{a} \right )^{4} \left ( \dfrac{\Sigma_{\mathrm{gas}} \pi a^2}{\MSun} \right )^{-1} \, \left ( \dfrac{M_{\mathrm{em}}}{\MSun} \right )^{-1} \, \left [ 1 + \frac{1}{4} \left ( \dfrac{e a}{z_s} \right )^{3} \right ]
	\label{eq:tau_e_type_I}
\end{equation}
\noindent where $\Sigma_{\mathrm{gas}} = \sqrt{\pi}z_s\rho_{\mathrm{gas}}$ is the local gas surface density, and $ z_s $ is the scale height of the disk.  The coefficients $ c_a $ and $ c_e $ control the strength of the radial drift and circularization of an embryo's orbit, respectively.  If the inclination damping timescale is not significantly shorter than the eccentricity damping timescale as \citet{2000MNRAS.315..823P} argue, then it will not contribute significantly to reaching the equilibrium state; thus for simplicity we assume $ \tau_{\mathrm{I,typeI}} = \tau_{\mathrm{e,typeI}}$.  Most agree that the choice $ c_e = 1.0 $ is reasonable; however there is less consensus about the exact value for the coefficient $ c_a $ though there is some agreement that $ c_a \lesssim 1.0 $ for the case of an isothermal equation of state for the gas.  However it has recently been shown by \citep{2009MNRAS.394.2283P} that the gas horseshoe torque can substantially modify the timescales in Eq.~\ref{eq:tau_a_type_I}, and even reverse the direction of migration, when the gas cooling timescale is greater than the horseshoe libration timescale.  However we shall limit our study to the ``classical" Type -- I formulae, and leave the inclusion of the horseshoe torque for a future study.

While we can include both aerodynamic and tidal gas interaction prescriptions in our $ N $-body code, we first discuss our simulations where we only consider aerodynamic gas drag in the next section.  We discuss the inclusion of Type -- I migration in \S\ref{type_I_effects}, which will be important for larger embryo masses (\eg $M_{\mathrm{em}} \gtrsim 1.0 \, \MEarth $).

\section{Simulations\label{simulations}}

We perform our numerical integrations using the SyMBA integrator \citep{1998AJ....116.2067D}, part of the SWIFT suite of packages.  SyMBA is a mixed-variable symplectic integrator based on the $ N $-body map of \citet{1991AJ....102.1528W}, which has been improved to accurately handle close encounters between particles.

In our simulations we consider two types of particles: massive bodies (\eg embryos) which mutually gravitate and can accrete other bodies, and less massive bodies (\eg planetesimals) which do not mutually gravitate or accrete.  Using a particle-mesh gravity solver, we are able to include the self-gravity of the planetesimals.  However for the disk masses considered in our simulations, the contribution from the planetesimal self-gravity is negligible at best, so we omit these calculations in all our simulations.  We have implemented the gas drag prescriptions as discussed in \S\ref{gas_effects}, where we assume aerodynamic gas drag and Type -- I migration applies to planetesimals and embryos, respectively.  Since we will be concentrating on embryos initially smaller than $ 1.0 \, \MEarth $, we have omitted the effects of Type -- I migration on the embryo for now.  We investigate the effects of Type -- I migration in \S\ref{type_I_effects}, particularly in conjunction with the planetesimal -- driven migration.

As a fiducial model, we assumed that the circumstellar disk corresponds to a power-law, where the volume density distribution of gaseous material:

\begin{equation}
	\rho_{\mathrm{gas}}(a, z) = f_g\rho_{\mathrm{gas,0}}\left(\dfrac{a}{\hbox{AU}} \right)^{-\alpha}e^{-[z/z_s(a)]^2}
	\label{eq:gas_volume_density}
\end{equation}
\noindent In a MMSN, $ \rho_{\mathrm{gas,0}} = 1.4 \times 10^{-9} \, \hbox{g cm}^{-3} $ at 1.0 AU and $ \alpha = 11/4 $ \citep{1981PThPS..70...35H}, while the disk scale height $ z_s(a) $ is given by a power-law:

\begin{equation}
	z_s(a) = z_{s,0}\left(\dfrac{a}{\hbox{AU}} \right)^{5/4}
	\label{eq:disk_scale_height}
\end{equation}
\noindent where $ z_{s,0} = 0.047 \, \hbox{AU} $ and $ f_g $ is a scaling factor of gaseous material in the disk, with $ f_g  = 1.0 $ in a MMSN.

The surface density distribution of solid material in a MMSN is also given by a power-law:

\begin{equation}
	\Sigma_{\mathrm{solid}}(a) = f_s f_{\mathrm{snow}}\Sigma_{\mathrm{solid,0}}\left(\dfrac{a}{\hbox{AU}} \right)^{-\beta}
	\label{eq:solid_surface_density}
\end{equation}
\noindent where $ \Sigma_{\mathrm{solid,0}} = 7.0 \, \hbox{g cm}^{-2} $ at 1.0 AU and $ \beta = \alpha - 5/4 $ assuming a gas to solid material ratio that does not vary radially, while

\begin{equation}
	f_{\mathrm{snow}} =
	\begin{cases}
		1.0 & \text{if $ a < a_{\mathrm{snow}} $} \\
		4.2 & \text{if $ a \geq a_{\mathrm{snow}} $}
	\end{cases}
\end{equation}
\noindent accounts for the enhancement of material beyond the snow-line, where the temperature of the circumstellar disk is cool enough for molecules (\eg H$ _2 $O, CO$ _2 $, CH$ _4 $, \etc) to condense into solids.  Typically $ a_{\mathrm{snow}} = 2.7 \, \hbox{AU} \, (L_{\star}/L_{\odot})^{1/2} $ \citep{1997ApJ...490..368C}, assuming a condensation temperature of 170 K.  The coefficient $ f_s $ is a scaling factor of solid material in the disk, with $ f_s = 1.0 $ in a MMSN.

Each of our simulations involve a single embryo on a circular orbit in the mid-plane of the disk, embedded in the centre of a $ 14 R_h $ wide annulus of planetesimals.  The planetesimals are given Rayleigh distributed \citep{1992Icar...96..107I} eccentricities and inclinations with dispersions $ \sigma_{e} = 2\sigma_{I} = 0.004 $, and we assume the density of each planetesimal is $ \rho_{\mathrm{plsml}} = 0.5 \, \hbox{g cm}^{-3} $.  All these simulations use $ 3.2 \times 10^4 $ equal-mass planetesimals, where the planetesimal mass is dependent on the disk model and the location of the embryo.

In most of our simulations we assume all of the planetesimals have the same radius $ R_{\mathrm{drag}} $, which is an independent parameter and does not depend on the mass or assumed density of the planetesimal.  Our planetesimals are actually tracer particles, each representing a large number of planetesimals of size $ R_{\mathrm{drag}} $ with similar orbits.  However we realize that using mono-dispersed planetesimals is not realistic, since we expect mutual collisions among the planetesimals to produce a size distribution.  It is not feasible to implement a self-consistent time dependent planetesimal size distribution at present, but we can implement a static planetesimal size distribution: $ dN/dR_{\mathrm{drag}} \propto R_{\mathrm{drag}}^{-\gamma} $ with $ 10^{-1} \, \hbox{km} \lesssim R_{\mathrm{drag}} \lesssim 10^{2} \, \hbox{km} $ and in \S\ref{size_distribution} we investigate the behaviour for a range of values: $ 2.0 \leq \gamma \leq 5.0 $.

\subsection{Effects of Aerodynamic Drag\label{effects_gas_drag}}

\begin{figure}[ht!]
	\begin{center}
	\includegraphics[width = \columnwidth]{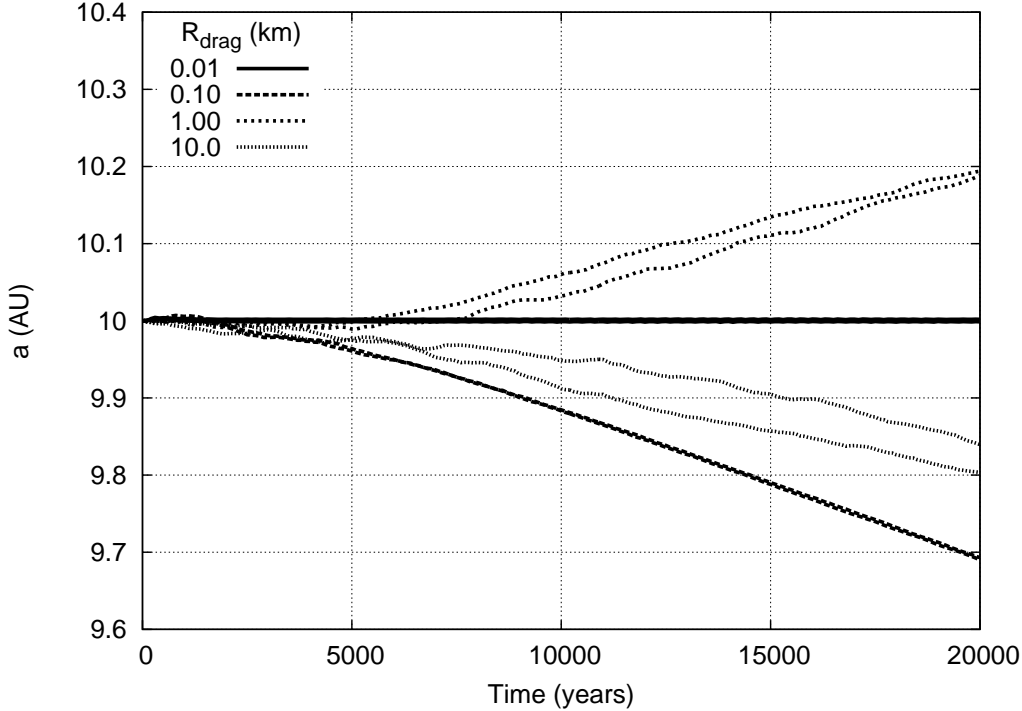}
	\caption[embryo_migration_regimes]{\label{fig:embryo_migration_regimes}
		Characteristic orbital evolution of a $ 1.0 \, \MEarth $ embryo started at 10.0 AU in a fiducial MMSN disk (\ie $ \alpha = 11/4 $ and $ f_g = f_s = 1.0 $) for different planetesimal radii $ R_{\mathrm{drag}} $.  Two different simulations are plotted for each $ R_{\mathrm{drag}} $, where the only difference is that the initial distribution of planetesimals were randomized.}
	\end{center}
\end{figure}

\begin{figure}[ht!]
	\begin{center}
	\includegraphics[width = \columnwidth]{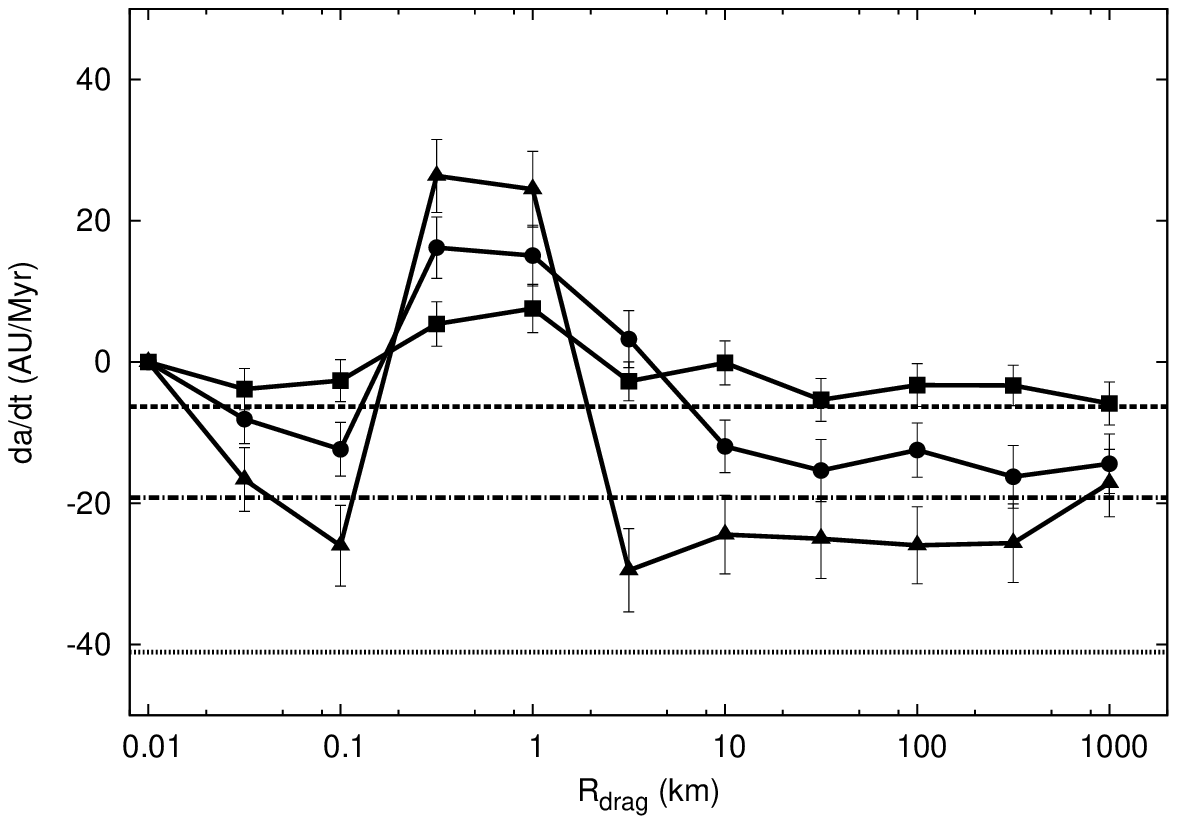}
	\caption[adot_rdrag_mem_025]{\label{fig:adot_rdrag_mem_025}
		Migration rate as a function of the planetesimal radius $ R_{\mathrm{drag}} $ for an $ 0.25 \, \MEarth $ embryo at 10.0 AU in a fiducial MMSN gas disk (\ie $ f_g = 1.0 $).  The squares, circles and triangles respond to simulations to disks with $ f_s = \{ 0.5, 1.0, 2.0 \} $, while the dashed, dot-dashed and dotted lines refer to the fiducial migration rate due to planetesimal scattering in Eq.~\ref{eq:adot_scatter}.  The migration rate, and its uncertainty, are measured from two realizations for each $ R_{\mathrm{drag}} $.  In these simulations, the value of $ C_D $ is computed based on the dynamical and physical properties of each planetesimal.}
	\end{center}
\end{figure}

\begin{figure}[ht!]
	\begin{center}
	\includegraphics[width = \columnwidth]{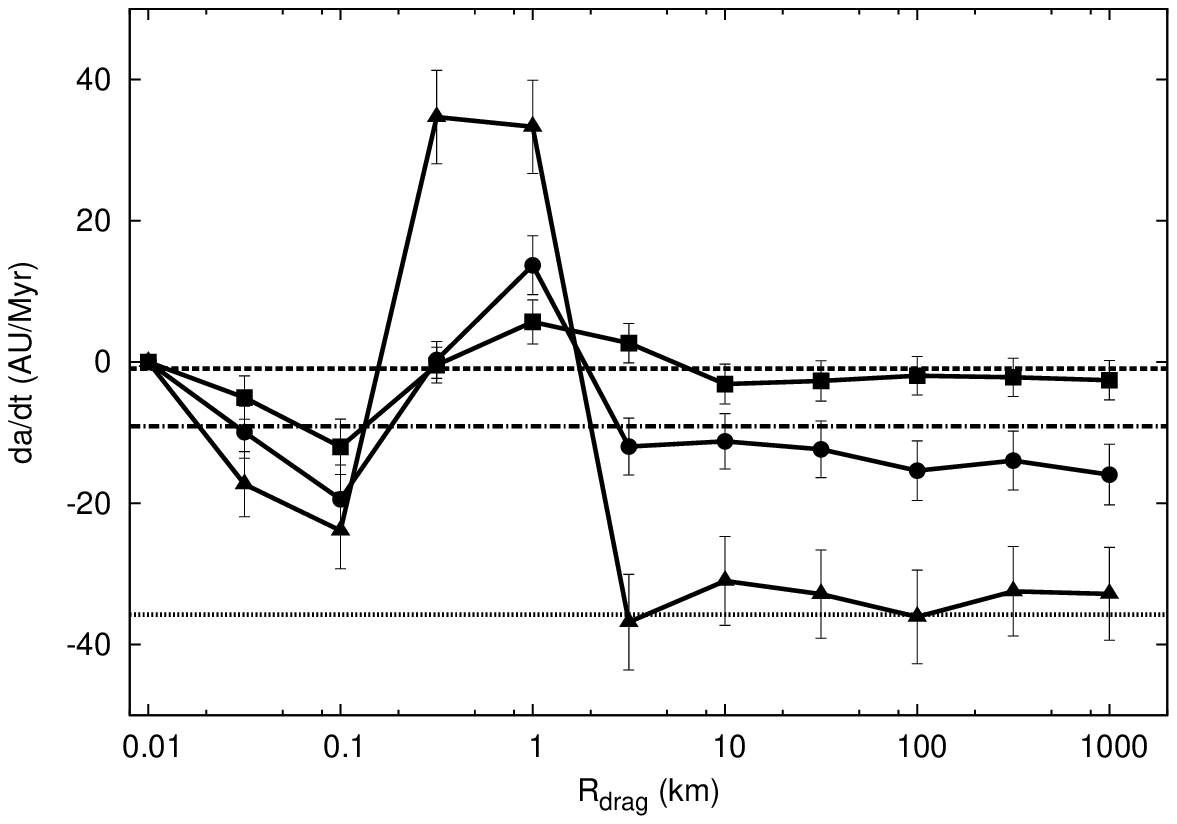}
	\caption[adot_rdrag_mem_10]{\label{fig:adot_rdrag_mem_10}
		Migration rate as a function of the planetesimal radius $ R_{\mathrm{drag}} $ for a $ 1.0 \, \MEarth $ embryo at 10.0 AU in a fiducial MMSN gas disk (\ie $ f_g = 1.0 $).  The squares, circles and triangles respond to simulations to disks with $ f_s = \{ 0.5, 1.0, 2.0 \} $, while the dashed, dot-dashed and dotted lines refer to the fiducial migration rate due to planetesimal scattering in Eq.~\ref{eq:adot_scatter}.  The migration rate, and its uncertainty, are measured from two realizations for each $ R_{\mathrm{drag}} $.  In these simulations, the value of $ C_D $ is computed based on the dynamical and physical properties of each planetesimal.}
	\end{center}
\end{figure}

We perform simulations for several combinations of parameters, namely: the mass of the embryo $ M_{\mathrm{em}} $, the orbital distance of the embryo $ a_{\mathrm{em}} $, the radius of the planetesimals $ R_{\mathrm{drag}} $, the gas volume density $ f_g $ and the solid surface density $ f_s $ relative to the fiducial MMSN model.  In particular we explored the following parameter values: $ M_{\mathrm{em}} =  \{0.25, 1.0 \} \, \MEarth $; $ a_{\mathrm{em}} = \{5.0, 10.0 \} \, \hbox{AU} $; $ f_g = \{0.5, 1.0, 2.0 \} $; $ f_s = \{0.5, 1.0, 2.0 \} $; and $ \log(R_{\mathrm{drag}}/\hbox{km}) $ from -2.0 to 3.0 in increments of 0.5.  For the two embryo locations at 5.0 AU and 10.0 AU, we integrate our simulations for $ 2 \times 10^4 \, \hbox{year} $ and $ 4 \times 10^4 \, \hbox{year} $, using time steps of 0.5 year and 1.25 year, respectively.  The results of these simulations are shown in Fig.~\ref{fig:adot_rdrag_mem_025} and Fig.~\ref{fig:adot_rdrag_mem_10}, which shows the rate of change of the embryo's orbital distance as a function of planetesimal radius.  The behaviour of the migration rate versus planetesimal radius is complicated, though we note four distinct migration regimes which are described in Fig.~\ref{fig:embryo_migration_regimes} for the case of a $ M_{\mathrm{em}} = 1.0 \, \MEarth $ embryo at $ a_{\mathrm{em}} = 10.0 \, \hbox{AU} $.

\subsubsection{$ R_{\mathrm{drag}} \lesssim 0.01 \, \hbox{km} $\label{rdrag_tiny}}

In this regime, the aerodynamic gas drag is so strong that it causes the semi-major axes and eccentricities of the planetesimals to decay very rapidly.  In fact, almost all the planetesimals tend to stream past the embryo before it has an opportunity to interact with them gravitationally.  The net result is the embryo does not migrate significantly at all.

\subsubsection{$ R_{\mathrm{drag}} \sim 0.1 \, \hbox{km} $\label{rdrag_small}}

In this regime, the aerodynamic gas drag is still strong, but the embryo is now able to gravitationally scatter some of the planetesimals.  Most of the planetesimals tend to become trapped in an exterior resonance with the embryo, but the aerodynamic gas drag is constantly damping the random motion of the planetesimals and causes their inward migration.  Since sufficient mass in planetesimals becomes trapped they are collectively able to push back against the embryo, and the net result is the embryo migrates inwards along with these trapped planetesimals.  Such behaviour was seen in the simulations of LTD10.

We can estimate the planetesimal radius that will be trapped in an exterior resonance, and the rate at which the embryo will migrate, based on the work of \citet{1993Icar..106..288K}.  They quantify the trapping condition through the drag parameter, $ K $:

\begin{equation}
	K = \dfrac{3\rho_{\mathrm{gas}}C_D}{8\rho_{\mathrm{plsml}}R_{\mathrm{drag}}}
	\label{eq:drag_parameter_kary}
\end{equation}
\noindent where they defined the critical drag parameter $ K_{\mathrm{trap}} $ to be where 50\% of the planetesimals become trapped in an exterior resonance, and $ \sim \hbox{50\%} $ stream past the embryo.  They determined the value of $ K_{\mathrm{trap}} $ numerically for several different planetary masses \citep[cf.][Table-III]{1993Icar..106..288K}.  If we fit the values found in this table from \citet{1993Icar..106..288K}, we obtain:

\begin{equation}
	K_{\mathrm{trap}}(M_{\mathrm{em}}) \simeq 1.9 \times 10^{-2} \, \hbox{AU}^{-1} \, \left ( \dfrac{M_{\mathrm{em}}}{\MEarth} \right )^{0.73}
	\label{eq:k_trap}
\end{equation}
 
Using this relation we can write $ R_{\mathrm{drag, trap}} $ for a MMSN disk model:

\begin{equation}
	R_{\mathrm{drag, trap}} \simeq 9.2 \, \hbox{km} \left ( \dfrac{C_D}{0.5} \right ) \left ( \dfrac{f_g}{1.0} \right ) \left ( \dfrac{0.5 \, \hbox{g cm}^{-3}}{\rho_{\mathrm{plsml}}} \right ) \left ( \dfrac{M_{\mathrm{em}}}{\MEarth} \right )^{-0.73} \left ( \dfrac{a}{\hbox{AU}} \right )^{-11/4}
	\label{eq:rdrag_trap}
\end{equation}

\begin{figure}[ht!]
	\begin{center}
	\includegraphics[width = \columnwidth]{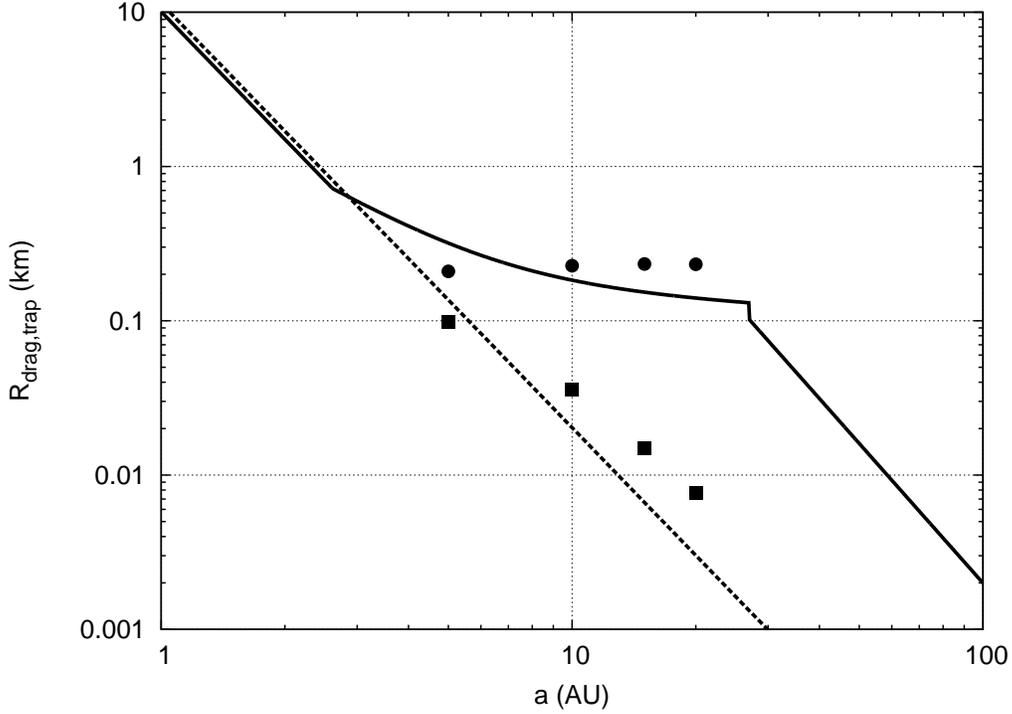}
	\caption[rdrag_trap]{\label{fig:rdrag_trap}
	The critical radius above which a planetesimal will become trapped in an exterior resonance with an embryo, as a function of the embryo's semi-major axis.  The transition radius is computed from the critical drag parameter $ K $ determined empirically by \citet{1993Icar..106..288K}, assuming either $ C_D = 0.5 $ (dashed line) or $ C_D $ computed based on the dynamical and physical properties of a typical planetesimal (solid line).  The points are results of simulations at different locations in the disk, squares for $ C_D = 0.5 $ and circles for computed $ C_D $.  In all the simulations the value $ K_{\mathrm{trap}} = 6.9 \times 10^{-3} \, \hbox{AU}^{-1} $ for a $ 0.25 \, \MEarth $ embryo is used, and the planetesimals are assumed to be massless.}
	\end{center}
\end{figure}

We test the predicted values in Eq.~\ref{eq:rdrag_trap} of $ R_{\mathrm{drag, trap}} $ by running simulations for the case of a $ 0.25 \, \MEarth $ embryo at four locations in a MMSN disk, using two choices of $ C_D $: fixed to 0.5 and computed based on the physical and dynamical properties of a typical planetesimal.  In the latter case, the value of $ R_{\mathrm{drag, trap}} $ in Eq.~\ref{eq:rdrag_trap} is solved iteratively since the value of $ C_D $ can depend on $ R_{\mathrm{drag}} $ in a nonlinear fashion.  To facilitate the determination of the fraction of trapped planetesimals, we assume that the planetesimals are massless in these simulations.  Illustrated in Fig.~\ref{fig:rdrag_trap} are the predicted values of $ R_{\mathrm{drag, trap}} $ as a function of distance in the disk, for both choices of $ C_D $.  Also plotted are results of the numerical experiments to verify $ R_{\mathrm{drag, trap}} $, and the points agree with the predicted values within a factor of three.  This discrepancy is more pronounced at larger distances, which may be a consequence of extrapolating $ R_{\mathrm{drag, trap}} $ beyond 5.0 AU where all the simulations of \citet{1993Icar..106..288K} were performed.

To estimate the rate at which the trapped planetesimals will push the embryo inwards, we compute the value of $ \dot{a}_{\mathrm{aero}} $ from Eq.~\ref{eq:adot_aero_drag} for mono-dispersed planetesimals of size $ R_{\mathrm{drag, trap}} $, assuming some RMS eccentricity $ e_{\mathrm{rms}} $ for the planetesimals.  However, the rate at which the ensemble of trapped planetesimals and the embryo will migrate inward will be reduced by a factor $ 1 + M_{\mathrm{em}}/M_{\mathrm{trap}} $, where $ M_{trap} $ is the mass of the planetesimals that remain trapped in an exterior resonance with the embryo.

\begin{equation}
	\dot{a}_{\mathrm{aero, trap}} = \dfrac{\dot{a}_{\mathrm{aero}}(a, e_{\mathrm{rms}}, R_{\mathrm{drag, trap}})}{1 + M_{\mathrm{em}}/M_{\mathrm{trap}}}
	\label{eq:adot_aero_trap}
\end{equation}

\begin{figure}[ht!]
	\begin{center}
	\includegraphics[width = \columnwidth]{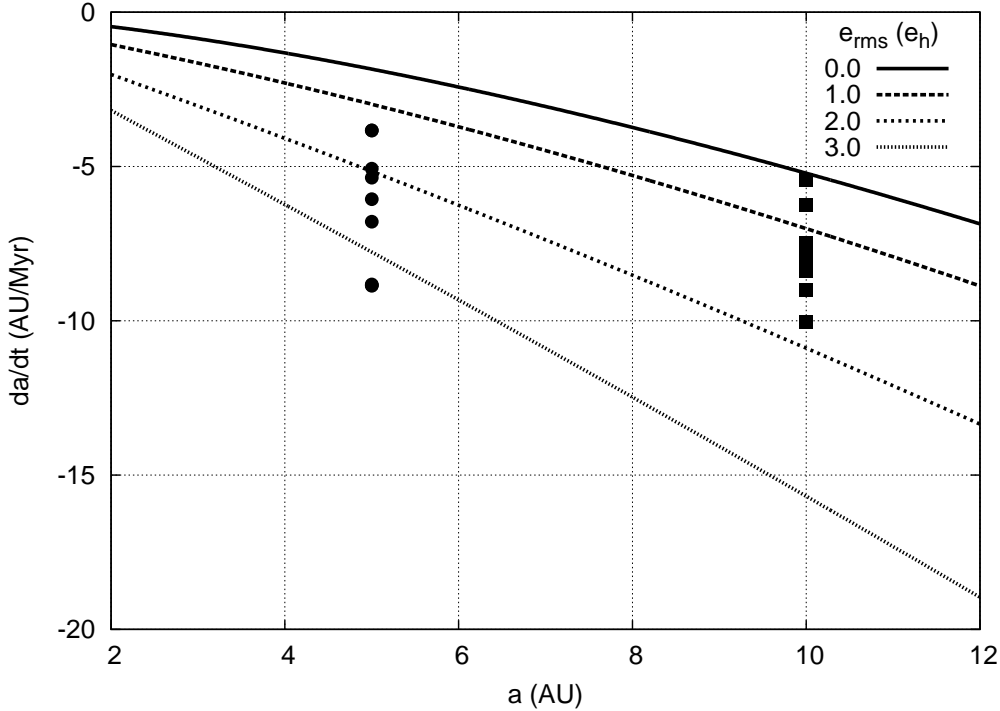}
	\caption[adot_aero_trap]{\label{fig:adot_aero_trap}
	The predicted migration rate $ \dot{a}_{\mathrm{aero, trap}} $ of a $ 0.25 \, \MEarth $ embryo as a function of the embryo's semi-major axis, that is being pushed by a population of trapped planetesimals of size $ R_{\mathrm{drag, trap}} $.  This calculation assumes that the planetesimal population trapped in an exterior resonance contains $ \sim 0.25 \, \MEarth $, which is consistent with the simulations.  Plotted are curves of $ \dot{a}_{\mathrm{aero, trap}} $ for different assumed RMS eccentricity $ e_{\mathrm{rms}} $ of the planetesimal population, in units of the Hill eccentricity $ e_h $.  The circles and squares correspond to the measured migration rate of the embryo from our simulations at two different locations, for planetesimal populations at or near $ R_{\mathrm{drag, trap}} $.}
	\end{center}
\end{figure}

For comparison with Eq.~\ref{eq:adot_aero_trap} we select the migration rates in Fig.~\ref{fig:adot_rdrag_mem_025}, at or near $ R_{\mathrm{drag, trap}} $ for two locations in a MMSN disk.  From these selected simulations, we estimate the typical mass of trapped planetesimals $ M_{trap} $ for use in Eq.~\ref{eq:adot_aero_trap}. Inspection of these simulations reveals that $ \sim 1 \, \hbox{--} \, 2 \, M_{\mathrm{em}}$ of planetesimals remain trapped in the exterior resonance.  Plotted in Fig.~\ref{fig:adot_aero_trap} is the measured migration rate of a $ 0.25 \, \MEarth $ embryo at two locations in a MMSN disk, for planetesimal populations at or near $ R_{\mathrm{drag, trap}} $.  Also plotted are the value of $ \dot{a}_{\mathrm{aero, trap}} $ from Eq.~\ref{eq:adot_aero_trap}, where the lines correspond to the assumed RMS eccentricity $ e_{\mathrm{rms}} $ for the planetesimals.  We note the good agreement between the measured and predicted migration rates for the range of $ e_{\mathrm{rms}} $ assumed.

\subsubsection{$ R_{\mathrm{drag}} \sim 1.0 \, \hbox{km} $\label{rdrag_medium}}

In this regime, the aerodynamic gas drag damps the random motions of planetesimals more slowly than that in \S\ref{rdrag_small}.  While the gas drag is diminished, it is still able to shift the relative populations of the scattering material on either side of the embryo.  This shift in the scattering populations is able to tip the imbalance in scattering events from outward to inward, and the net effect is the embryo tends to migrate outward in response.  This will be discussed in greater detail in \S\ref{toy_model}.  The speed of the outward migration of the embryo is also comparable to the inward migration in the gas-free case, which seems to indicate that planetesimal -- driven migration is the driving mechanism in the outward migration case.  The maximum outward migration rate for $ R_{\mathrm{drag}} \sim 1.0 \, \hbox{km} $ appears to be reproduced in most of the combinations of the parameters that we examined, with minor variations.

\subsubsection{$ R_{\mathrm{drag}} \gtrsim 10.0 \, \hbox{km} $\label{rdrag_large}}

In this regime, the aerodynamic gas drag becomes increasingly irrelevant.  This is evident as the migration rate asymptotically approaches the gas-free migration rate, which in most cases agrees favourably with the estimate in Eq.~\ref{eq:adot_scatter} from \citet{2009Icar..199..197K}.  These are plotted as dashed, dot-dashed and dotted horizontal lines in Fig.~\ref{fig:adot_rdrag_mem_10} and Fig.~\ref{fig:adot_rdrag_mem_025}.  However, the results for the $ 0.25 \, \MEarth $ simulations for $ f_s = 2.0 $ do not agree as favourably with Eq.~\ref{eq:adot_scatter}.  Since the other sets of simulations do agree, it is worth noting that the migration rate in Eq.~\ref{eq:adot_scatter} from \citet{2009Icar..199..197K} is an estimate that is accurate only to within a factor of two \citep[cf.][Fig.~6]{2009Icar..199..197K}, which is consistent with our results.

We can also estimate the largest radius of mono-dispersed planetesimals that results in outward migration of the embryo, using our understanding of gas-free planetesimal -- driven migration and aerodynamic drag.  We recall that in the gas-free case, planetesimal scattering preferentially induces inward migration of the embryo.  This arises from the slight imbalance of angular momentum transferred to the embryo by scattering planetesimals from either side of its orbit, leaving the embryo with a dearth of angular momentum and hence causing it to migrate inwards \citep{2009Icar..199..197K}.  With gas present, the aerodynamic gas drag will cause the orbits of planetesimals to circularize, thereby removing the planetesimals from the embryo's encounter zone.

So we propose that the two relevant timescales are the timescale for the aerodynamic gas drag to damp a planetesimal's eccentricity $ e $ by a factor of order itself (\ie $ \tau_{\mathrm{e,aero}} \equiv \left | e/\dot{e} \right |_{\mathrm{aero}} $), and the migration timescale of the embryo due to planetesimal scattering (\ie $ \tau_{\mathrm{a,sca}} \equiv \left | a/\dot{a} \right |_{\mathrm{sca}} $).  Equating the two timescales determines a rough scaling relation for the transition between outward and inward embryo migration.

If we substitute Eq.~\ref{eq:gas_volume_density} for the gas volume density for a MMSN in the mid-plane of the disk and the Keplerian velocity, the aerodynamic gas drag timescale Eq.~\ref{eq:tau_aero} can be written:

\begin{equation}
	\tau_{\mathrm{aero}} \simeq 2.0 \, \hbox{year} \left ( \dfrac{1.0}{f_g} \right ) \left ( \dfrac{R_{\mathrm{drag}}}{\hbox{km}} \right )\left ( \dfrac{a}{\hbox{AU}} \right )^{13/4}
	\label{eq:tau_aero_numerical}
\end{equation}
\noindent where we have substituted in for the values of $ \rho_{\mathrm{plsml}} = 0.5 \, \hbox{g cm}^{-3} $ and $ C_D = 0.5 $.

Using Eq.~\ref{eq:edot_aero_drag}, Eq.~\ref{eq:tau_aero_numerical} and assuming $ I = e/2 $, we can write:

\begin{equation}
	\tau_{\mathrm{e,aero}} = \dfrac{2.3 \times 10^2 \, \hbox{year}}{f(a,e)} \left ( \dfrac{1.0}{f_g} \right ) \left ( \dfrac{R_{\mathrm{drag}}}{\hbox{km}} \right ) \left ( \dfrac{a}{\hbox{AU}} \right )^{13/4}
	\label{eq:tau_e_aero}
\end{equation}
\noindent where $ f(a,e) = \sqrt{\eta_0^2(a/\hbox{AU}) +7.5 \times 10^{-5}(M_{\mathrm{em}}/\MEarth)^{2/3}e_h^2} $ with $ \eta_0 = 1.95 \times 10^{-3} \, \hbox{AU}^{-1/2}$ and $ e_h \equiv e/\xi_h $ is the Hill eccentricity.

Using Eq.~\ref{eq:adot_scatter}, we can define a migration timescale due to planetesimal scattering:

\begin{equation}
	\tau_{\mathrm{a,sca}} \simeq 4.8 \times 10^4 \, \hbox{year} \left ( \dfrac{1.0}{f_s} \right ) \left ( \dfrac{a}{\hbox{AU}} \right )
	\label{eq:tau_a_sca_numerical}
\end{equation}

Comparing the timescales given in Eq.~\ref{eq:tau_e_aero} and Eq.~\ref{eq:tau_a_sca_numerical}, we define the maximum planetesimal radius (\ie $ R_{\mathrm{drag}} \leq R_{\mathrm{drag,trans}} $) that results in outward migration of an embryo through the criterion $ \tau_{\mathrm{e,aero}} \leq \tau_{\mathrm{a,sca}} $:

\begin{equation}
	R_{\mathrm{drag,trans}} \simeq 2.1 \times 10^2 \, \hbox{km} \, f(a,e) \left (\dfrac{f_g}{f_s} \right ) \left ( \dfrac{a}{\hbox{AU}} \right )^{-9/4}
	\label{eq:rdrag_trans}
\end{equation}

\begin{figure}[ht!]
	\begin{center}
	\includegraphics[width = \columnwidth]{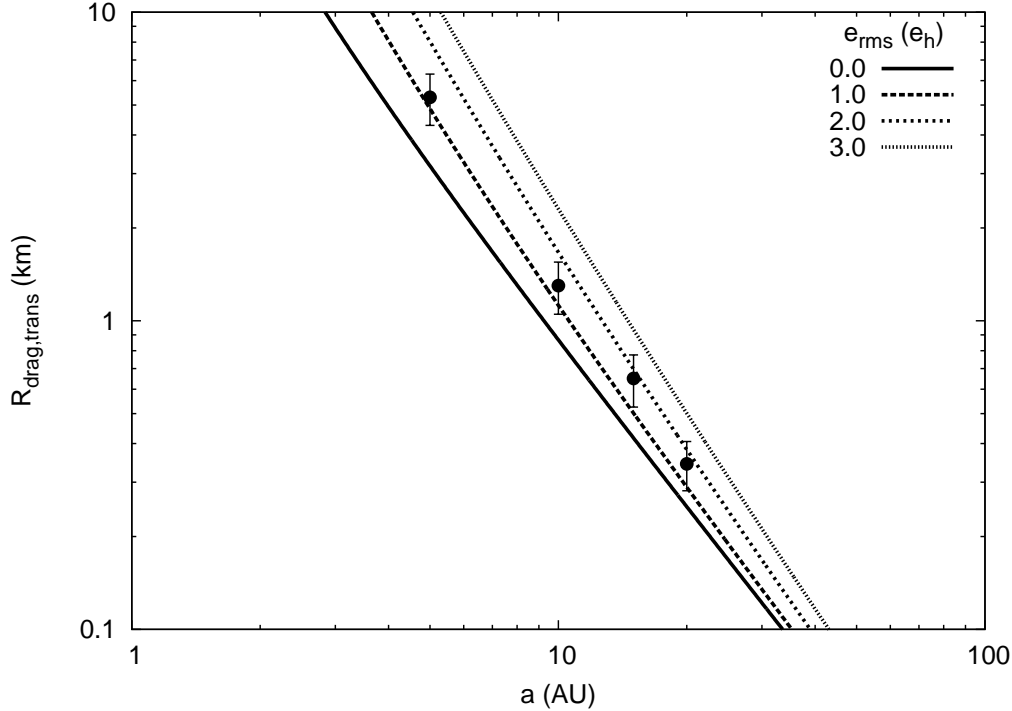}
	\caption[rdrag_trans]{\label{fig:rdrag_trans}
		Comparison of analytical estimate of the maximum $ R_{\mathrm{drag}} $ that results in outward migration from Eq.~\ref{eq:rdrag_trans} in a fiducial MMSN disk, with results from simulations.  Plotted are curves of $ R_{\mathrm{drag,trans}} $ for different assumed RMS eccentricity $ e_{\mathrm{rms}} $ of the planetesimal population, in units of the Hill eccentricity $ e_h $.  The points and error bars are estimates of $ R_{\mathrm{drag,trans}} $, extrapolated from five sets of 16 identical simulations at each location in the disk. The value of the drag coefficient is fixed at $ C_D = 0.5 $ in these simulations.}
	\end{center}
\end{figure}

In Fig.~\ref{fig:rdrag_trans} we have plotted as a function of $ a $ the values of $ R_{\mathrm{drag,trans}} $ from Eq.~\ref{eq:rdrag_trans}, for four different $ e_{\mathrm{rms}}$.  Also plotted are the results from a series of simulations that were produced to determine the transition location. We see that this simple scaling relation is able to predict the transition to outward embryo migration.

In the next section we create a more comprehensive toy model that addresses the different migration regimes discussed in this section: inward migration, outward migration and gap clearing.  The model will make predictions for the transitions between these different migration regimes.

\section{Toy Model\label{toy_model}}

In this section we consider a toy model which displays the main features of the dynamical regimes described above.  Indeed, with representative choices of the parameters it gives rough quantitative predictions for the planetesimal size for which outward migration is most likely to be triggered for a single embryo embedded in a dynamically cold planetesimal disk when aerodynamic drag effects are included.

We begin by recalling some key results described in \citet{2009Icar..199..197K} for the gas-free case:
\begin{enumerate}
	\item As noted in section \S\ref{grav_effects}, for sufficiently low mass embryos, planetesimal -- driven migration is self-sustaining in either direction at a rate given by Eq.~\ref{eq:adot_scatter}.
	\item The steady-state migration develops from an instability caused by a positive feedback loop in which the migration initially accelerates at a rate proportional to the migration rate \citep[cf.][for analogous result in Type -- III migration in gas disks]{2003ApJ...588..494M}.
	\item Unless there is a very strong positive disk density gradient in the outward direction, inward migration is the typical outcome.
	\item The trigger for the inward migration arises from a slight asymmetry between the rates that angular momentum is transferred to the embryo by outward versus inward scattering of planetesimals.
\end{enumerate}

The last point is demonstrated using results from the circular restricted three-body problem, in which the planetesimals have negligible total mass compared to the embryo.  Starting with an initial dynamically cold population, \citet{2009Icar..199..197K} monitored the populations in the so-called ``encounter" zones on either side of the embryo, which are the two regions of semi-major axes extending from roughly 1.0 -- 3.5 $ R_h $ interior and exterior to the planet. Because of the existence of a constant of the motion known as the Jacobi integral, only particles in the encounter zones can come within the Hill sphere of the planet and have a strong scattering (or be accreted). In \citet{2009Icar..199..197K} it is shown that there is an asymmetry in the transfer rates between the two reservoirs that leads to a build-up of particles exterior to the embryo. When the planetesimals have non-zero mass, the build-up of exterior particles corresponds to an outward flux of planetesimal angular momentum which, by conservation of angular momentum, causes the embryo to move inward.  As fresh planetesimals interior to the embryo enter the feeding zone, the transfer asymmetry removes more angular momentum from the embryo, triggering a positive feedback \ie an instability.  Although the entire process is a complicated series of scatterings together with a slow diffusion in the planetesimal eccentricities, our simple model is parameterized by a single transfer timescale for each reservoir, together with a single estimate of the specific angular momentum exchanged with the embryo as a particle is transferred from one side to the other.

Thus our toy model has two reservoirs - one interior to the planet containing total time dependent mass $ M_{\mathrm{int}}(t) $ and one exterior with total mass $ M_{\mathrm{ext}}(t) $.  We define $ \Delta\tilde{J} $ to be the gain of specific angular momentum of a particle scattered from the interior to the exterior reservoir. In the model we assume the planet of mass $ M_{\mathrm{em}} $ is fixed on a circular orbit of semi-major axis $ a_{\mathrm{em}} $ and infer the gain/loss in its angular momentum $ J_{\mathrm{em}} $ by equating it to the negative of the loss/gain of angular momentum of the scattering planetesimals. In the gas free case, then, the relevant equations for the toy model are:

\begin{subequations}
	\label{eq:toy_model}
	\begin{align}
		\dot{M}_{\mathrm{int}} &= - \dfrac{M_{\mathrm{int}}}{\tau_{\mathrm{int}}} + \dfrac{M_{\mathrm{ext}}}{\tau_{\mathrm{ext}}} \label{eq:mdot_int} \\
		\dot{M}_{\mathrm{ext}} &= - \dfrac{M_{\mathrm{ext}}}{\tau_{\mathrm{ext}}} + \dfrac{M_{\mathrm{int}}}{\tau_{\mathrm{int}}} \label{eq:mdot_ext} \\
		\dot{J}_{\mathrm{em}} &= \left ( \dfrac{M_{\mathrm{int}}}{\tau_{\mathrm{int}}} - \dfrac{M_{\mathrm{ext}}}{\tau_{\mathrm{ext}}} \right ) \Delta\tilde{J} \label{eq:jdot_embryo}
	\end{align}
\end{subequations}
\noindent where $ \tau_{\mathrm{int}} $ and $ \tau_{\mathrm{ext}} $ are the timescales for a planetesimal to be scattered out of the interior or exterior reservoir, respectively.

The choice of the parameters in the model may be determined from restricted three-body scattering experiments \citep[cf.][Fig.~1 and Fig.~2 and accompanying discussion]{2009Icar..199..197K}. The initial asymmetry in the inner and outer encounter zones means $ M_{\mathrm{ext}}(0) = M_{\mathrm{int}}(0)(1 + 5.5\xi_h) $. The timescale for scattering a particle from the inner zone is the product of the timescale for each encounter (\ie the synodic period) times the probability per encounter that such a scattering occurs.  From Fig.~1 of \citet{2009Icar..199..197K} we find $ \tau_{\mathrm{int}} = 6.7/\xi_h $ orbital periods to be representative of a particle which starts near the middle of the encounter zone. The value of $ \tau_{\mathrm{ext}} $ can be inferred from Fig.~2 of \citet{2009Icar..199..197K}, together with the fact that the asymptotic solution of the equations above is $ \tau_{\mathrm{ext}}/\tau_{\mathrm{int}} = M_{\mathrm{ext}}/M_{\mathrm{int}} $: we find $ \tau_{\mathrm{ext}}/\tau_{\mathrm{int}} = (1 + 12\xi_h) $.

Finally, the value of $ \Delta\tilde{J} $ can be estimated by noting that the conservation of the Jacobi constant during an outward scattering event typically transforms a particle with eccentricity $ e $ and apocentre near $ a_{\mathrm{em}} - R_h $ to one with pericentre near $ a_{\mathrm{em}} + R_h $ and comparable eccentricity. It is easy to show that for small eccentricity, this results in a change of specific angular momentum equal to $ (\xi_h + e)\tilde{J}_{\mathrm{em}} $, where $ \tilde{J}_{\mathrm{em}} = \sqrt{G\MSun a_{\mathrm{em}}} $ is the specific angular momentum of a circular orbit at the planet's location.

The eccentricities of particles in the encounter zones are typically a few $ \xi_h $.  However, \citet{2007PhDT........21K} found that particles within a given encounter zone diffusing to larger eccentricities (due to successive small scatterings) also transferred angular momentum to the embryo.  He found that the particles in the exterior reservoir evolved to higher eccentricities than those in the interior reservoir.  Thus the net effect of diffusion in the two reservoirs results in a net gain in angular momentum of the planetesimals.  Indeed he found the net angular momentum transferred by particles jumping from one reservoir to the other to be comparable in magnitude and sign to the net angular momentum transferred by diffusion within the reservoirs. This is incorporated in the model by using a somewhat larger value of the angular transfer rate than would be the case for particles that are only scattered between wings. We find a representative choice to be $ \Delta\tilde{J} = 8\xi_h\tilde{J}_{\mathrm{em}}$.  

\begin{figure}[ht!]
	\begin{center}
	\includegraphics[width = \columnwidth]{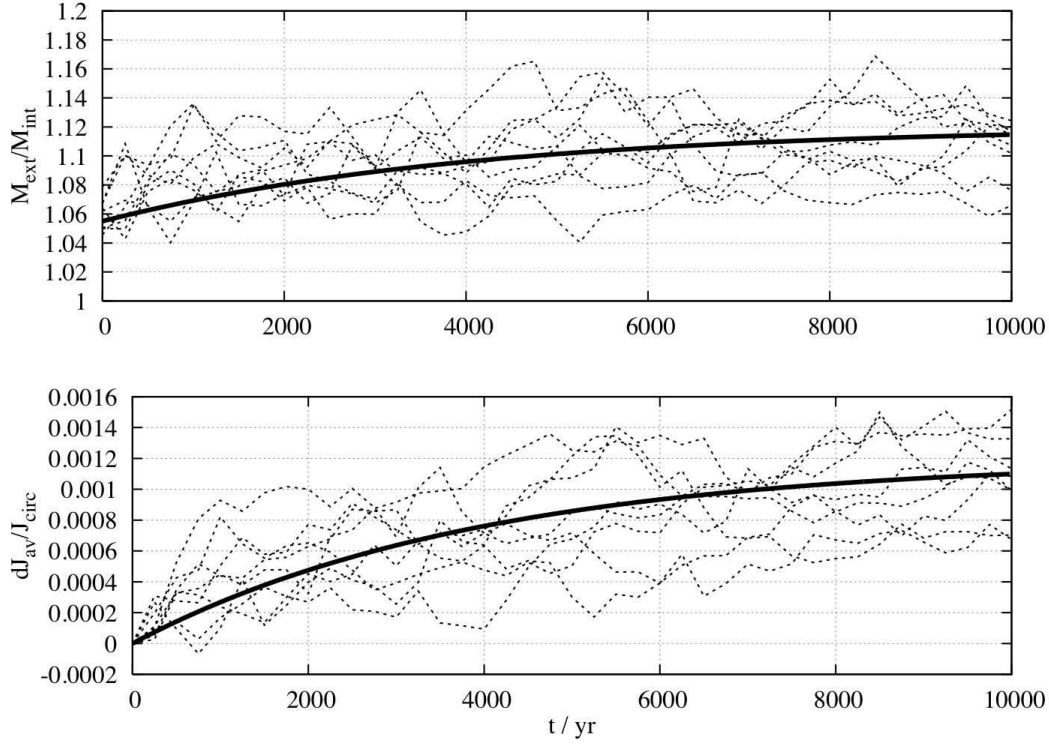}
	\caption[dndj]{\label{fig:dndj}
		Upper panel: the evolution with time of the ratio of masses in the two reservoirs on either side of a $ 1.0  \, \MEarth $ planet at 5.0 AU. Lower panel: the average change in angular momentum of the planetesimals versus time.  The solid curves are the prediction of the toy model discussed in the text. The dashed curves are the results of eight $ N $-body simulations of the same system using different random seeds for the initial planetesimal positions.}
	\end{center}
\end{figure}

Fig.~\ref{fig:dndj} shows the behaviour of our toy model for the case of a $ 1.0  \, \MEarth $ planet on a circular orbit at 5.0 AU embedded in a sea of massless planetesimals. The upper solid curve shows the ratio predicted by Eq.~\ref{eq:toy_model} for the number of particles in the outer zone versus those in the inner encounter zone. The bottom solid curve shows the predicted average amount of angular momentum transferred per particle in units of the specific circular angular momentum.  In both plots the results of direct $ N $-body simulations are shown for comparison. It can be seen that the agreement is very good, as it is for several other cases which we ran with different mass planets.

Let us now consider the net angular momentum transferred to the embryo when aerodynamic drag acts on the planetesimals.  Since the aerodynamic drag tends to circularize orbits, planetesimals scattered outward onto eccentric orbits tend to have their perihelion distance drawn from near the Hill sphere of the embryo to locations further away from the strong scattering region.  Indeed if the eccentricity damping timescale is shorter than the timescale for another strong scattering, the particle may evolve into a near-circular orbit several Hill radii from the embryo where it can be trapped in a mean-motion resonance exterior to the embryo and no longer have strong scatterings.  On the other hand, planetesimals scattered inward will have both their aphelion distances and semi-major axes drawn inward.  For rapid damping, planetesimals scattered inward may then be dragged inward and not have another strong scattering with the embryo.  The aerodynamic drag can be thought of as producing a ``sink" for planetesimals from the two reservoirs.  Moreover since the aerodynamic drag depends on the gas density, which itself has a strong radial dependence, there will be significant differences in the timescales for the draining of planetesimals from the interior versus exterior reservoir.  In our simple toy model we thus add two terms associated with the removal of particles due to aerodynamic drag and drag timescales $ \tau_{\mathrm{drag,int}} $ and $ \tau_{\mathrm{drag,ext}} $, for each reservoir respectively.  The equations for our toy model now become:

\begin{subequations}
	\label{eq:toy_model_drag}
	\begin{align}
		\dot{M}_{\mathrm{int}} &= \dot{M}_{\mathrm{sca}} - \dfrac{M_{\mathrm{int}}}{\tau_{\mathrm{drag,int}}}  \label{eq:mdot_drag_in} \\
		\dot{M}_{\mathrm{ext}} &= - \dot{M}_{\mathrm{sca}} - \dfrac{M_{\mathrm{ext}}}{\tau_{\mathrm{drag,ext}}}  \label{eq:mdot_drag_out} \\
		\dot{J}_{\mathrm{em}} &=- \dot{M}_{\mathrm{sca}}\Delta\tilde{J} \label{eq:jdot_drag_embryo}
	\end{align}
\end{subequations}
\noindent where $ \dot{M}_{\mathrm{sca}} \equiv - M_{\mathrm{int}}/\tau_{\mathrm{int}} + M_{\mathrm{ext}}/\tau_{\mathrm{ext}} $ is the net mass loss rate due to planetesimal scattering for the interior reservoir.

Although analytic solutions to Eq.~\ref{eq:toy_model_drag} exist, for the purposes of computing specific angular momentum transferred to the planet simple numerical integrations illustrate the qualitative nature of the solutions, and these can indeed be quantitatively compared to the results of the full $ N $-body simulations. For illustrative purposes, we adopt the gas disk density profile of LTD10, in which the gas volume density is $ 3.4 \times 10^{-9} \, \hbox{g cm}^{-3} $ at 1.0 AU and $ \alpha = 9/4 $. For simplicity, in both the toy model and the $ N $-body simulations we fix the drag coefficient $ C_D = 0.5 $.  We insert a planet with mass $ M_{\mathrm{em}} = 1.0 \, \MEarth $ on a circular orbit at 5.0 AU in a planetesimal disk with surface density such that the initial mass in planetesimals in the inner encounter zone is $M_{\mathrm{int}}(0) = 2.0 \, M_{\mathrm{em}} $.  A typical particle in the scattered wings will have $ e \approx 3\xi_h $ and using the definitions above we find for a planetesimal with drag radius $ R_{\mathrm{drag}} $ that $ \tau_{\mathrm{drag,int}} = 674 \, \hbox{year} \, (R_{\mathrm{drag}}/\hbox{km}) $ and $ \tau_{\mathrm{drag,ext}} = 1382 \, \hbox{year} \, (R_{\mathrm{drag}}/\hbox{km}) $.

Experience with the $ N $-body simulations shows that once the planet has migrated a distance $ \sim R_h/3 $ in one direction the instability has been initiated and the planet continues to migrate in that direction. Thus we define:

\begin{equation} 
   \Delta J_{\mathrm{crit}} = M_{\mathrm{em}} \left [ \sqrt{G\MSun \left ( a_{\mathrm{em}} + \frac{1}{3}R_h \right ) } - \tilde{J}_{\mathrm{em}} \right ] \approx \frac{1}{6}M_{\mathrm{em}}\tilde{J}_{\mathrm{em}}\xi_h
   \label{eq:dj_crit}
\end{equation}
\noindent as the critical amount of angular momentum required to initiate self-sustaining migration.

\begin{figure}[ht!]
	\begin{center}
	\includegraphics[width = \columnwidth]{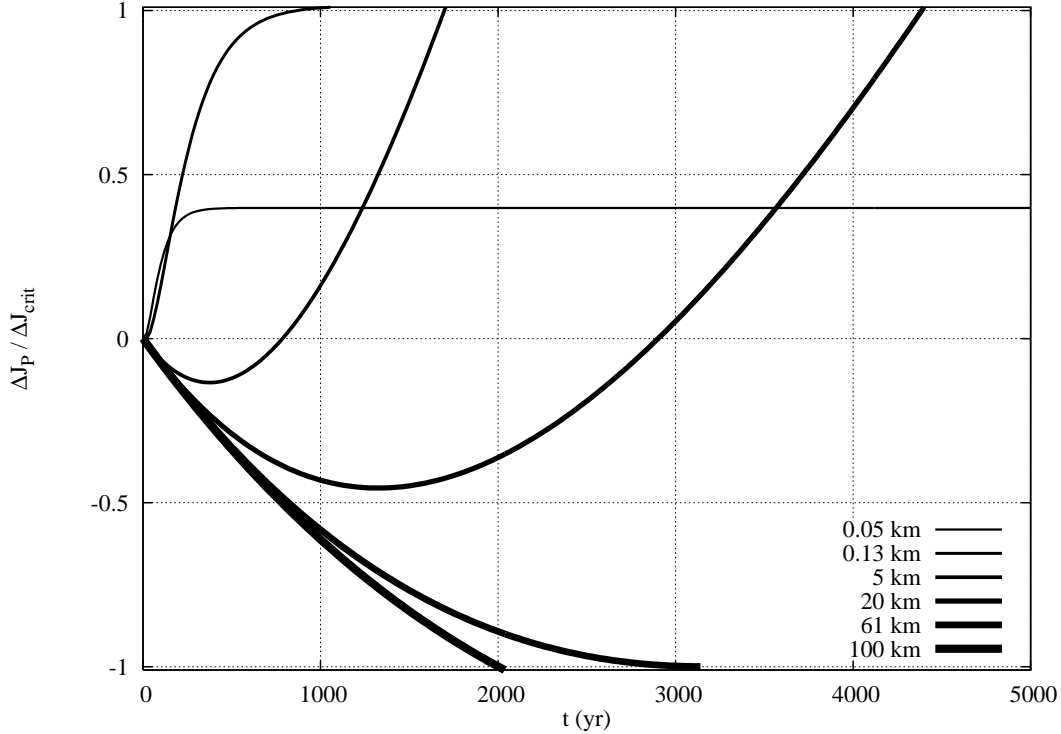}
	\caption[toy_all]{\label{fig:toy_all}
		The time evolution of the angular momentum transferred to the planet for the toy model parameters discussed in the text. Thicker curves correspond to larger planetesimals, as shown in the legend.}
	\end{center}
\end{figure}

\begin{figure}[ht!]
	\begin{center}
	\includegraphics[width = \columnwidth]{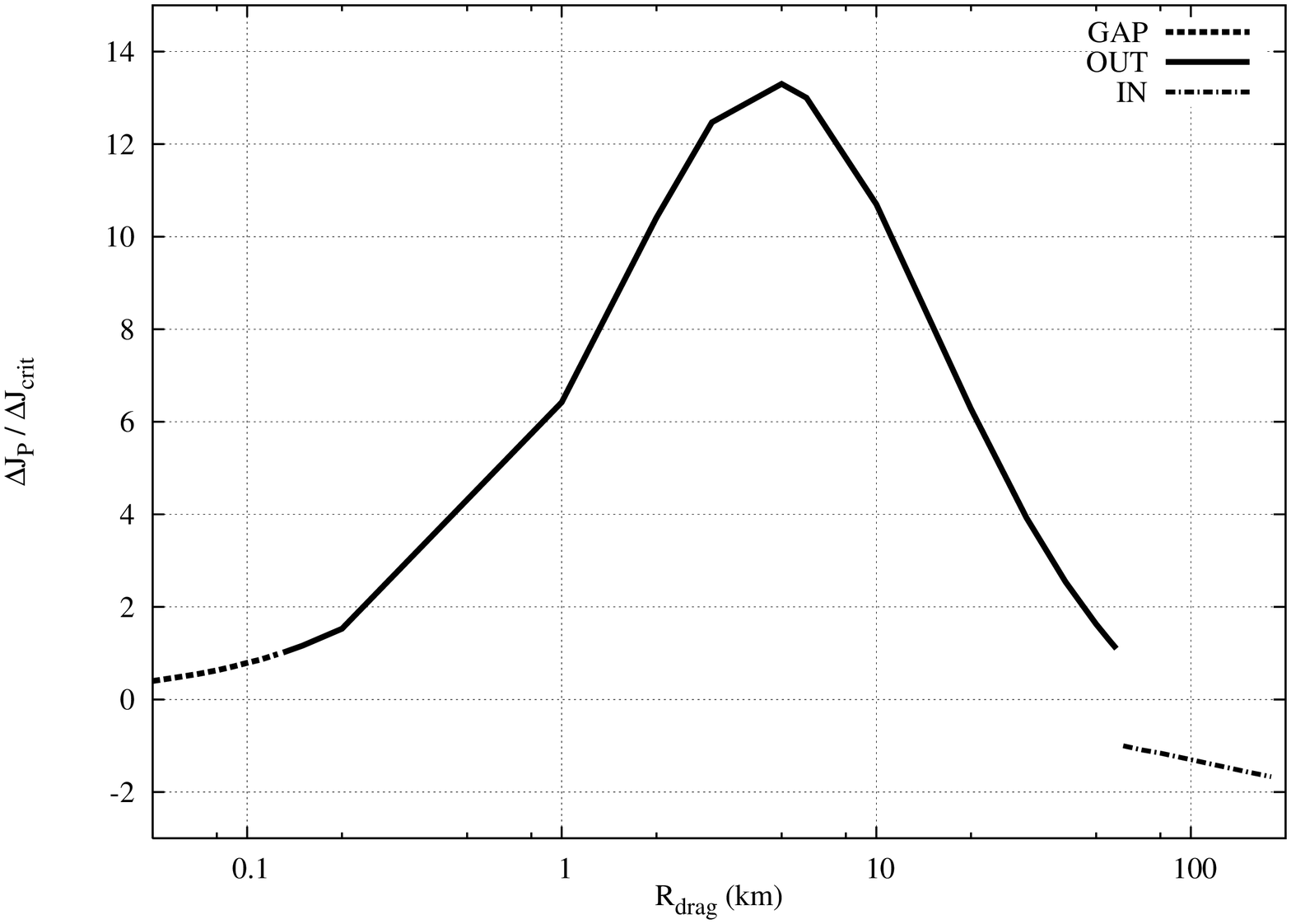}
	\caption[toy_dj_rdrag]{\label{fig:toy_dj_rdrag}
		The first extremum in Fig.~\ref{fig:toy_all} of the angular momentum transferred to the planet versus $ R_{\mathrm{drag}} $ for the toy model.}
	\end{center}
\end{figure}

Plotted in Fig.~\ref{fig:toy_all} are the predictions of the toy model for the time evolution of the angular momentum transferred to the planet for the disk parameters given above and for several values of the planetesimal drag radius.  For each value of $ R_{\mathrm{drag}} $ the curve is terminated if the magnitude of the transfer equals $ \Delta J_{\mathrm{crit}} $ since at that time self-sustaining migration (either inward or outward) is assumed to ensue.  An alternative way to view the predictions of the toy model is shown in Fig.~\ref{fig:toy_dj_rdrag}, in which the first extremum in the  angular momentum transferred is shown as a function of $ R_{\mathrm{drag}} $. There it can be seen that for $ R_{\mathrm{drag}} < 0.6 \, \hbox{km} $ a gap is predicted because insufficient angular momentum is transferred to trigger migration, and since in that case the encounter zones become evacuated, a gap in the planetesimal disk forms around the planet.  For $ R_{\mathrm{drag}} > 20 \, \hbox{km} $, sufficiently negative angular momentum is transferred in the early stages to trigger inward migration.  For values in between these drag radii outward migration is predicted.  The discontinuity near $ R_{\mathrm{drag}} = 20 \, \hbox{km} $ corresponds to a transition between solutions which ``bottom-out" at $ \Delta J_{\mathrm{em}} < -\Delta J_{\mathrm{crit}} $ and those which do not.

\begin{figure}[ht!]
	\begin{center}
	\includegraphics[width = \columnwidth]{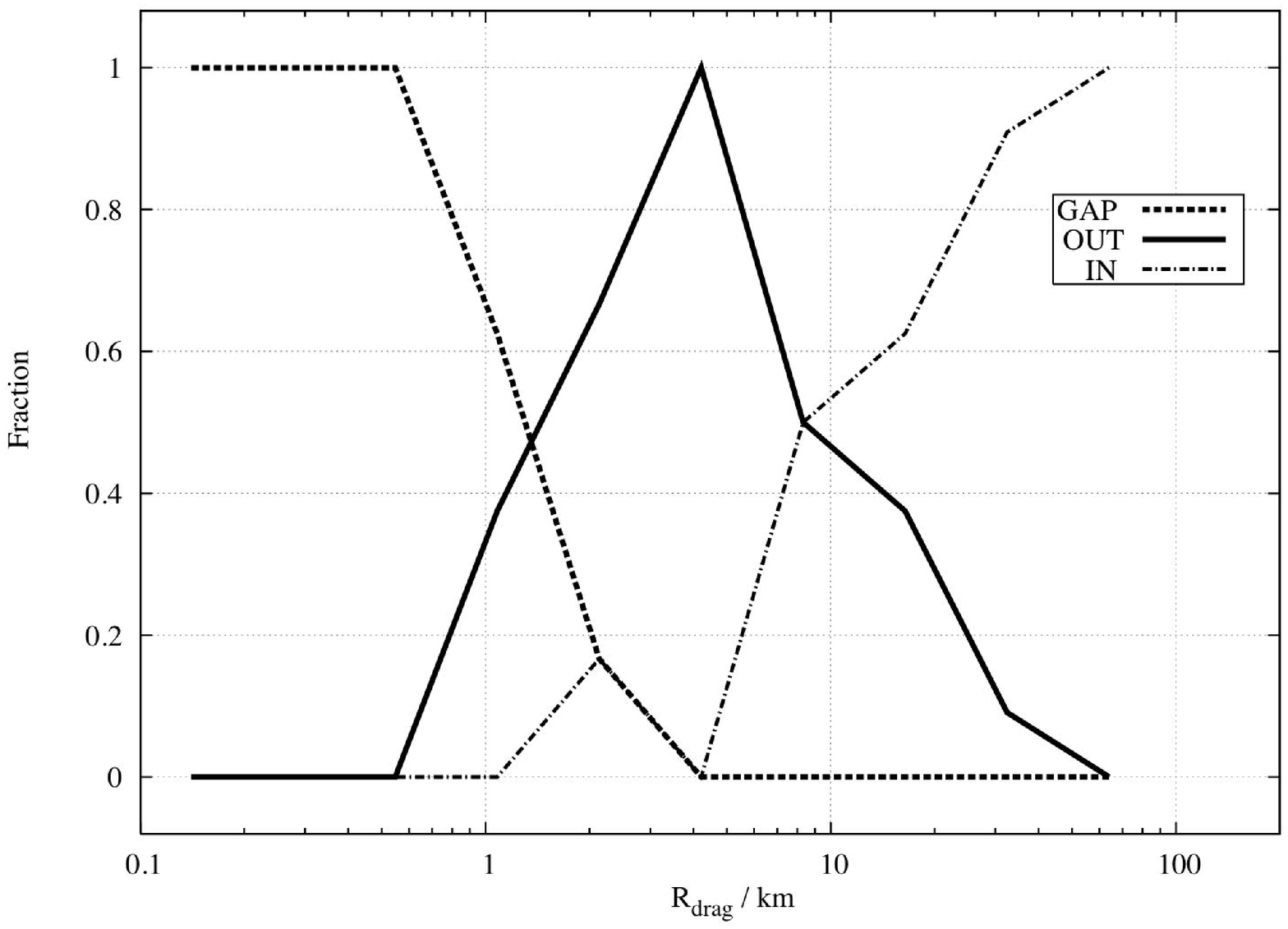}
	\caption[frsum]{\label{fig:frsum}
		Summary of results from multiple $ N $-body simulations of the evolution of a $ 1.0 \, \MEarth $ embryo starting at 5.0 AU for the disk parameters described in the text. Three curves are shown: for each value of planetesimal drag radius $ R_{\mathrm{drag}} $ the solid curve shows the fraction migrating out, the dot-dash curve shows the fraction migrating in and the dashed curve displays the fraction migrating in.}
	\end{center}
\end{figure}

In full $ N $-body simulations, individual scattering is of course stochastic and the fairly subtle asymmetries involved therefore lead to stochastic outcomes. To compare with the toy model for the same disk parameters we have run numerous $ N $-body experiments and binned the outcomes into 3 possible scenarios: outward migration, inward migration and planetesimal gap formation. The simulations involve a $ 1.0 \, \MEarth $ embryo at 5.0 AU, with a population of $ 3.0 \times 10^4 $ dynamically cold (\ie $ \sigma_e = 2\sigma_I = 10^{-3} $) planetesimals distributed from 4.0 AU to 6.0 AU, each with a different $ R_{drag} $ value.  Approximately 16 $ N $-body simulations were performed, in each of 10 logarithmically spaced bins in $ R_{\mathrm{drag}} $.  Each simulation involved a $ 1.0  \, \MEarth $ embryo at 5.0 AU, embedded in a sea of massless planetesimals. We note that Fig.~\ref{fig:toy_dj_rdrag} and Fig.~\ref{fig:frsum} agree with each other, at least on a qualitative level. While the toy model is able to capture the physics underlying embryo migration, it fails to account for the minutia of the planetesimals' dynamical behaviour.  So while our toy model can reasonably predict the planetesimal radius which most often leads to outward embryo migration, it also predicts values for $ R_{\mathrm{drag, trap}} $ and $ R_{\mathrm{drag, trans}} $ that are almost an order of magnitude too small and too large, respectively.

We can extrapolate the results in Fig.~\ref{fig:frsum} as a function of $ M_{em} $ and $ a_{em} $ via the transitional radii defined by Eq.~\ref{eq:rdrag_trap} and Eq.~\ref{eq:rdrag_trans}. If we assume $ C_D = 0.5 $ for simplicity then for fixed $ M_{em} $, $ R_{\mathrm{drag, trap}} $ and $ R_{\mathrm{drag, trans}} $ will both decrease with increasing $ a_{em} $.  However, since $ R_{\mathrm{drag, trap}} $ decreases faster than $ R_{\mathrm{drag, trans}} $, the width of the regime for outward embryo migration will increase with $ a_{em} $.  In the case for fixed $ a_{em} $, $ R_{\mathrm{drag, trap}} $ will decrease, while $ R_{\mathrm{drag, trans}} $ will increase with $ M_{em} $.  So the width of the regime for outward embryo migration will expand with $ M_{em} $, albeit differentially.  While an extrapolation of the results in Fig.~\ref{fig:frsum} could also be performed with a $ C_D $ computed based on the dynamical and physical properties of a typical planetesimal, the nonlinear nature of $ C_D $ does not permit for simple scaling relations.  In addition, extensive numerical simulations to confirm these predictions would be computationally expensive.
  
\begin{figure}[ht!]
	\begin{center}
	\includegraphics[width = \columnwidth]{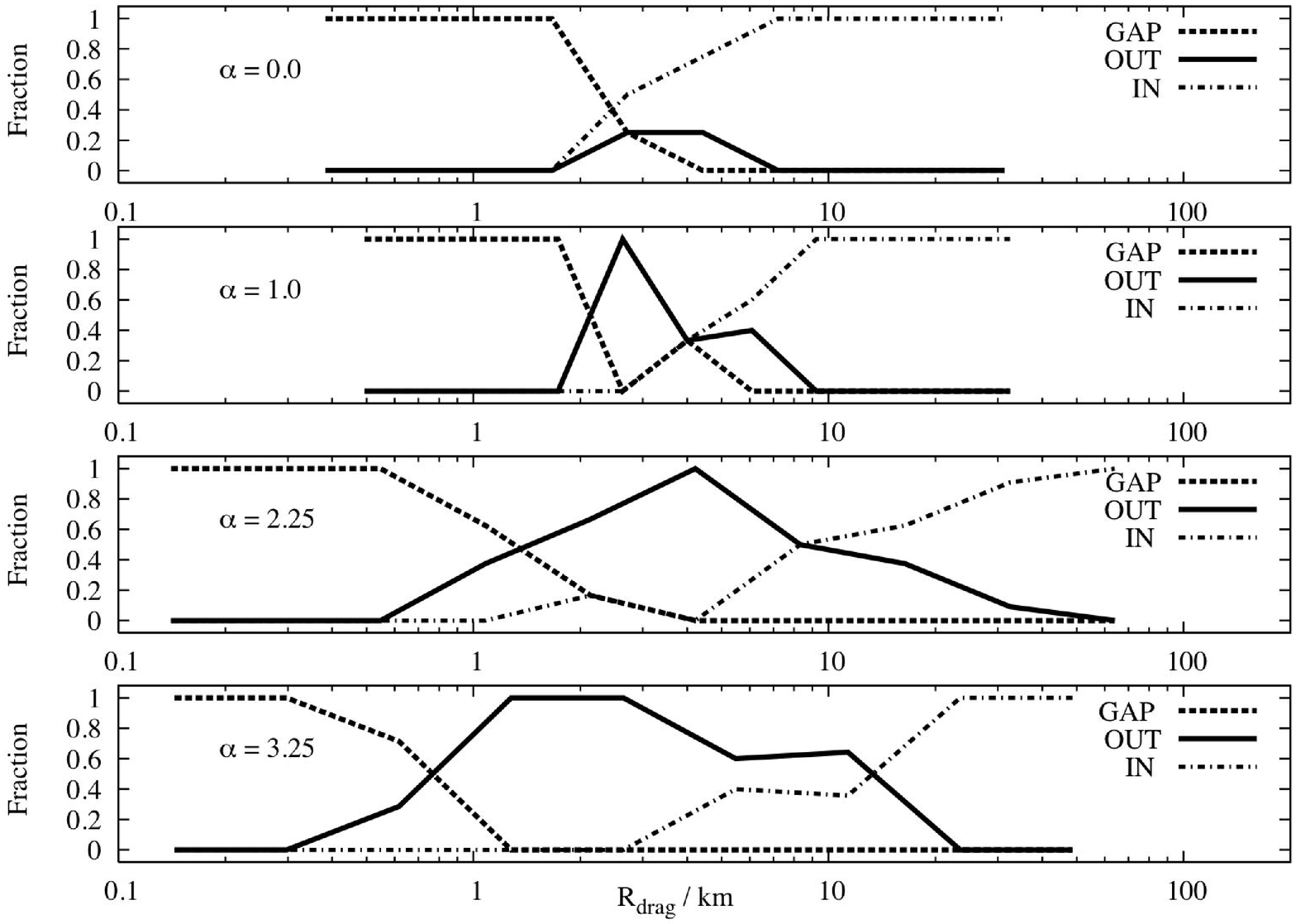}
	\caption[frall]{\label{fig:frall}
		Similar to Figure~\ref{fig:frsum} but for $ N $-body simulations of a $ 1.0 \, \MEarth $ mass planet at $ a = 5.0 \, \hbox{AU} $ in which the gas density is $ 9.0 \times 10^{-11} \, \hbox{g cm}^{-3} \, (a/5.0 \, \hbox{AU})^{-\alpha} $, $ C_D = 0.5 $, and
 $ \eta(a) = 6.0 \times 10^{-4} \, (\alpha + \frac{1}{2}) (a/\hbox{AU})^{1/2} $.  From top to bottom, the results shown are for $ \alpha = \{ 0.0, 1.0, 2.25, 3.25 \}$, respectively.}
	\end{center}
\end{figure}

To demonstrate the effect of changing the slope of the gas density profile, Fig.~\ref{fig:frall} shows the results of $ N $-body simulations of a $1.0 \, \MEarth $ mass planet at 5.0 AU in which the gas volume density is $ 9.0 \times 10^{-11} \, \hbox{g cm}^{-3} $ at 5.0 AU, $ C_D = 0.5 $ and $ \eta $ is determined as in \S\ref{aero_gas_drag}.  By varying $ \alpha $ and keeping the gas to solid ratio fixed (\eg $ \beta = \alpha - 5/4 $) we can see the important effects on outward migration that a large gradient in gas density creates.  Recall that the typical semi-major axes of particles in the exterior reservoir, are somewhat larger than in the interior reservoir.  Since the gas drag timescale is inversely proportional to the local gas density, a strong gradient in gas density thus leads to a large disparity in the gas drag timescale between the two reservoirs.  Thus particles in the inner reservoir are differentially depleted more rapidly when the gas density gradient is large.  The results of the toy model (not shown) are in good qualitative agreement with the $ N $-body simulations.

\section{Size Distribution\label{size_distribution}}

Up to this point we have considered only mono-dispersed planetesimals in our simulations, but as was noted earlier this is not very realistic.  While incorporating a self-consistent time dependent planetesimal size distribution is not feasible at present, we did want to address concerns about the combined effect of a size distribution on embryo migration.  To that end, we implement a static planetesimal size distribution: $ dN/dR_{\mathrm{drag}} \propto R_{\mathrm{drag}}^{-\gamma} $ for planetesimal sizes $ 0.0625 \, \hbox{km} \leq R_{\mathrm{drag}} \leq 32.0 \, \hbox{km} $, and we investigate the behaviour of the embryo migration for a range of power-law exponents: $ 2 \leq \gamma \leq 5 $.  Of special note are two values of the power-law exponent, namely $ \gamma = 7/2 $ and $ \gamma = 4 $.  In the case of $ \gamma = 7/2 $, this value corresponds to the size distribution resulting from a collisional cascade \citep{1969JGR....74.2531D}, while the value of $ \gamma = 4 $ corresponds to the case where the mass in the size distribution is logarithmically equally-spaced.

\begin{figure}[ht!]
	\begin{center}
	\includegraphics[width = \columnwidth]{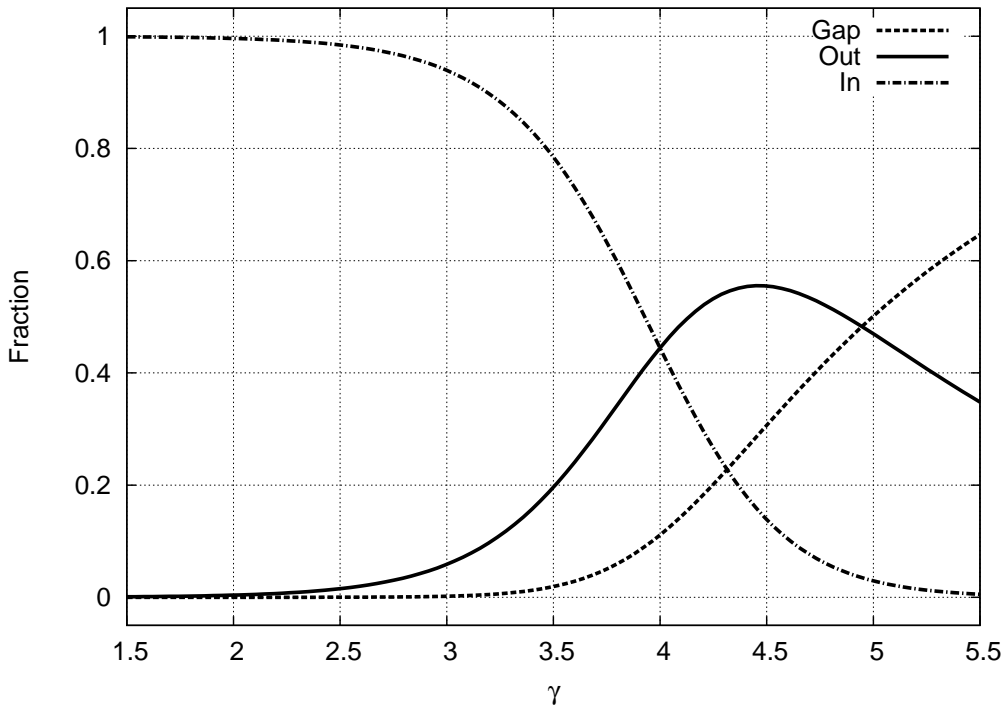}
	\caption[fsweet_gamma]{\label{fig:fsweet_gamma}
		The fraction of the total mass contained in planetesimals that lead to one of the migration outcomes: inward migration, outward embryo migration or gap clearing, as a function of the exponent $ \gamma $ for the planetesimal size distribution.  For a $ 0.25 \, \MEarth $ embryo at 10.0 AU in a fiducial MMSN disk, the transitions between these three different outcomes occurs at: $ \sim 0.125 \, \hbox{km}$ and $ \sim 2.0 \, \hbox{km} $.  This assumes a range $0.0625 \, \hbox{km} \leq R_{\mathrm{drag}} \leq 32.0 \, \hbox{km} $ for the planetesimal size distribution.}
	\end{center}
\end{figure}

It stands to reason that if a significant fraction of the mass in the planetesimal size distribution preferentially causes the embryo to migrate in a certain direction, then that population of planetesimals will dictate the outcome of embryo migration.  We can estimate the range of $ \gamma $ that will give rise to the different migration outcomes (cf. \S\ref{effects_gas_drag}) by computing the fraction of the total mass those $ R_{\mathrm{drag}} $ will contribute that give rise to the migration outcomes (cf. Fig.~\ref{fig:fsweet_gamma}).  For an embryo of mass $ 0.25 \, \MEarth $ at 10.0 AU in a fiducial MMSN disk, the gap clearing regime occurs for $ R_{\mathrm{drag}} < 0.125 \, \hbox{km} $.  For the case of outward migration, that occurs for $ 0.125 \, \hbox{km} \lesssim R_{\mathrm{drag}} \lesssim 2.0 \, \hbox{km} $, and finally for inward migration occurs for $ R_{\mathrm{drag}} > 2.0 \, \hbox{km} $.  The fraction of the total mass that these different regimes contribute are plotted in Fig.~\ref{fig:fsweet_gamma}.  What this plot indicates is that there will be a smooth transition between the different migration regimes, but since the migration outcome is bimodal, stochastic effects will be important.

\begin{figure}[ht!]
	\begin{center}
	\includegraphics[width = \columnwidth]{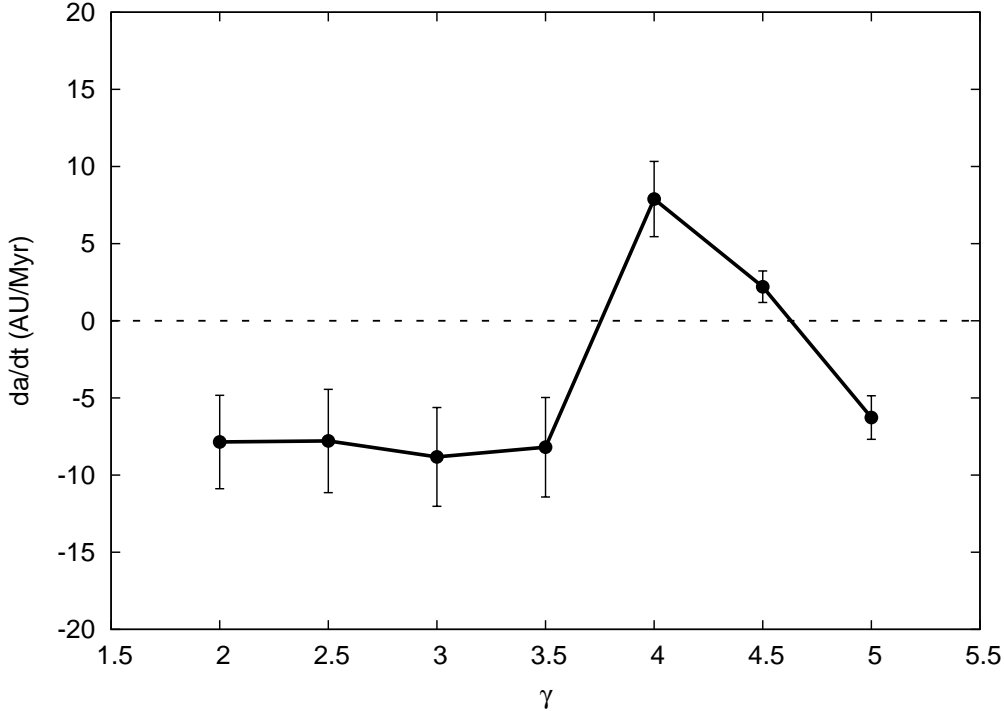}
	\caption[adot_gamma]{\label{fig:adot_gamma}
		Migration rate for an $ 0.25 \, \MEarth $ embryo at 10.0 AU in a fiducial MMSN disk, as function of the exponent $ \gamma $ of the planetesimal size distribution.  The average of 16 identical simulations were used to determine the migration rate and its uncertainty, where only the initial planetesimal distribution were randomized.  The value of $ C_D $ is computed based on the dynamical and physical properties of each planetesimal in these simulations, and a range of planetesimal radii: $ 0.0625 \, \hbox{km} \leq R_{\mathrm{drag}} \leq 32.0 \, \hbox{km} $ were used.}
	\end{center}
\end{figure}

\begin{figure}[ht!]
	\begin{center}
	\includegraphics[width = \columnwidth]{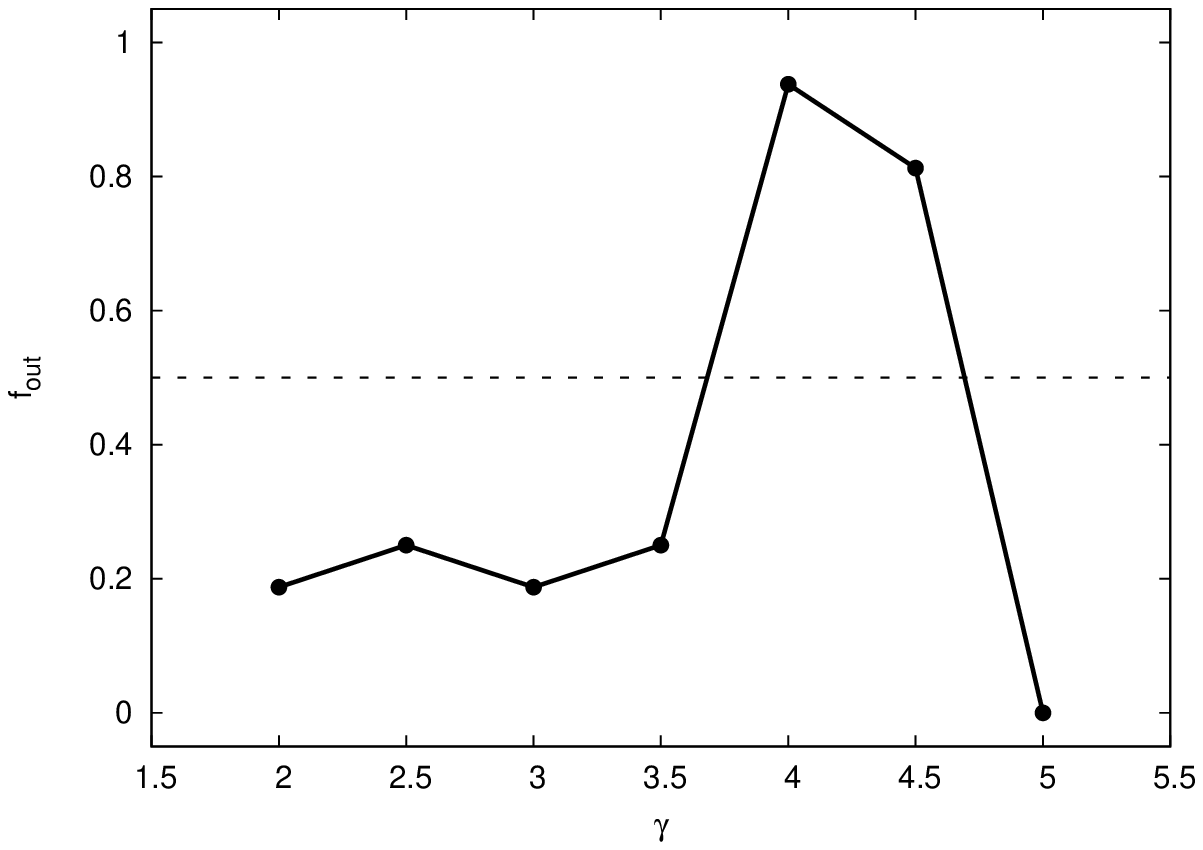}
	\caption[fout_gamma]{\label{fig:fout_gamma}
		Fraction of simulations that resulted in outward migration as a function of the exponent $ \gamma $ of the planetesimal size distribution, for a $ 0.25 \, \MEarth $ embryo at 10.0 AU in a fiducial MMSN disk.  Sets of 16 identical simulations were used to determine the outward fraction, where only the initial planetesimal distribution were randomized.  In these simulations.  The value of $ C_D $ was computed based on the dynamical and physical properties of each planetesimal in these simulations, and a range of planetesimal radii: $0.0625 \, \hbox{km} \leq R_{\mathrm{drag}} \leq 32.0 \, \hbox{km} $ were used.  The $ f_{\mathrm{out}} = 0.5 $ threshold (dashed line) was used to determine range of $ \gamma $ that results in outward migration.}
	\end{center}
\end{figure}

To determine what occurs when we include a size distribution, we perform 16 sets of simulations for a $ 0.25 \, \MEarth $ embryo at 10.0 AU in a fiducial MMSN disk, each for seven values of $ \gamma $.  The average migration rate for the embryo from these simulations are plotted in Fig.~\ref{fig:adot_gamma} as a function of $ \gamma $.  We see that for $ 4.0 \leq \gamma \leq 4.5 $, the average embryo migration rate is outwards while for all other values of $ \gamma $ the embryo migrated inwards.  In Fig.~\ref{fig:fout_gamma} we plot the fraction of the 16 simulations that result in outward migration, and we note that for $ \gamma < 3.5 $ there are still $ \sim 20 \hbox{--} 40\% $ of the simulations that result in outward migration.  This is despite the fact that these planetesimal size distributions heavily favour large $ R_{\mathrm{drag}} $, and consequently inward migration.  If we now define the 50\% fraction as the threshold defining outward embryo migration, then the range of $ \gamma $ that results in outward migration is: $ 3.7 \lesssim \gamma_{\mathrm{out}} \lesssim 4.7 $.  For this range of $ \gamma $, the majority of the mass in size spectrum are in those planetesimals that cause the embryo to migrate outward.  So for disk models with a different range of planetesimal radii causing outward migration of the embryo, then we would expect the range of $ \gamma $ to shift to smaller or larger values depending if the planetesimal radii ``sweet spot" makes the size spectrum less or more top-heavy.

So we can see that for a reasonable range of $ \gamma $ the planetesimal size distribution results in outward embryo migration, but even outside this range there is still a small chance that the embryo will migrate outwards.  Next we look at extending our analysis to include Type -- I migration, and see how it competes with planetesimal -- driven migration and aerodynamic gas drag.

\section{Effects of Type -- I\label{type_I_effects}}
\begin{figure}[ht!]
	\begin{center}
	\includegraphics[width = \columnwidth]{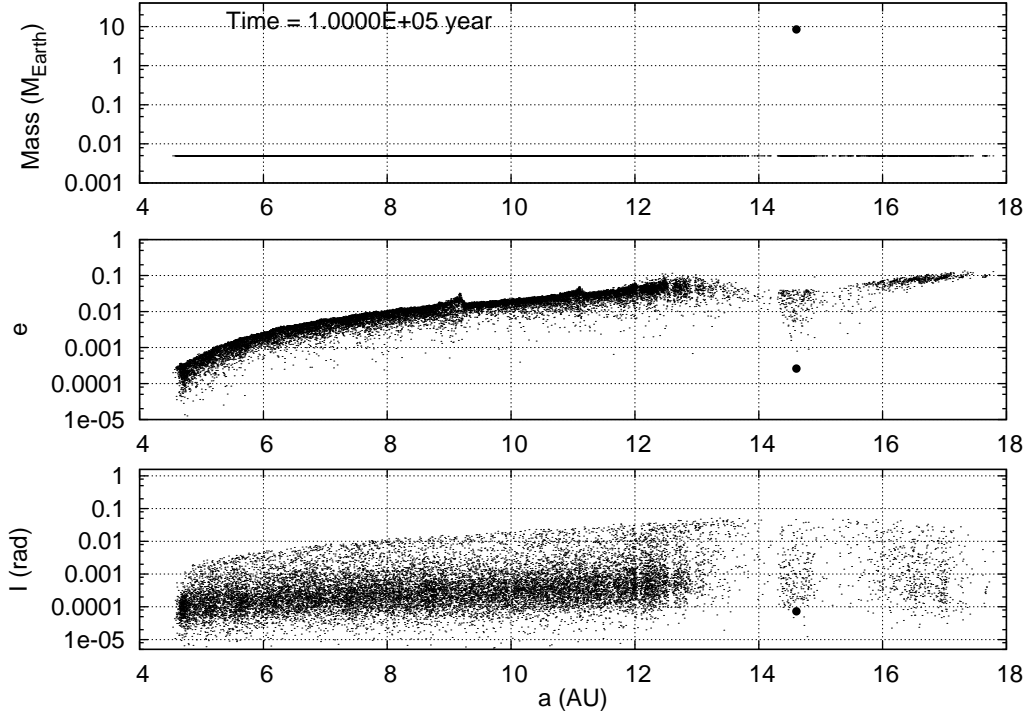}
	\caption[outward_migration_growth]{\label{fig:outward_migration_growth}
		Simulation of a $ 0.25 \, \MEarth $ embryo at 5.0 AU in a disk with $ f_g = f_s = \sqrt{5} $, $ \alpha = 9/4 $ and 1.0 km size planetesimals.  After $ 10^5 \, \hbox{yrs} $, the embryo migrated outwards to $ \sim 14.5 \, \hbox{AU} $ and attained a mass of $ \sim 8.5 \, \MEarth $.}
	\end{center}
\end{figure}

We now turn our attention to Type -- I migration, and address its ability to deposit embryos and giant planetary cores on the central star.  We could argue that this tendency implies that the efficiency parameter $ c_a $ as it appears in Eq.~\ref{eq:tau_a_type_I}, must be much less than unity for large embryo masses.  While decreasing $ c_a $ would solve the issue with embryo deposition, it cannot be justified in general except where the physical conditions permit it (\eg depleted gaseous disks, turbulence, \etc).  As we have shown in \S\ref{effects_gas_drag}, planetesimal -- driven migration can lead to rapid inward and outward embryo migration.  Furthermore, simulations with outward migration also result in the rapid growth of the embryo, particularly in dynamically cold disks.  An illustration is provided in Fig.~\ref{fig:outward_migration_growth}, where a $ 0.25 \, \MEarth $ embryo is placed at 5.0 AU in disk similar to the example used in \S\ref{toy_model} with $ f_g = f_s = \sqrt{5} $ and $ \alpha = 9/4 $ (LTD10) with a planetesimal size $ R_{\mathrm{drag}} = 1.0 \, \hbox{km} $.  At the end of the simulation, the embryo migrates outwards to $ \sim 14.5 \, \hbox{AU} $ and grows to a mass of $ \sim 8.5 \, \MEarth $.  This example suggests that including a population of cold, collision fragments, especially if planetary envelopes are included \citep{2003A&A...410..711I} could aid in the rapid formation of a core.

Under conditions for outward planetesimal -- driven embryo migration and growth, Type -- I migration will act to counteract this outward migration.  It stands to reason with the embryo accreting planetesimals, there will be a point where Type -- I migration will dominate planetesimal -- driven migration.  To assess the relative importance of these two competing migration mechanisms, we start by asking when would the timescale for Type -- I migration $ \tau_{\mathrm{a,typeI}} $ be shorter than the timescale for planetesimal -- driven migration $ \tau_{\mathrm{a,sca}} $.  The formula for these two timescales are given below, and in the case of $ \tau_{\mathrm{a,typeI}} $ we assume $ e \lesssim 2\xi_h $.

\begin{equation}
	\tau_{\mathrm{a,typeI}} = \dfrac{P_{\mathrm{orb}}}{2 \pi c_a} \left ( \dfrac{z_s}{a} \right )^2 \left [ \dfrac{\Sigma_{\mathrm{gas}}(a)\pi a^2}{\MSun} \right ]^{-1} \left ( \dfrac{M_{\mathrm{em}}}{\MSun} \right )^{-1}
	\label{eq:tau_a_type_I_e_zero}
\end{equation}

\begin{equation}
	\tau_{\mathrm{a,sca}} = \dfrac{P_{\mathrm{orb}}}{2} \left [ \dfrac{\Sigma_{\mathrm{solid}}(a)\pi a^2}{\MSun} \right ]^{-1} \left[1 + \dfrac{1}{5} \left ( \dfrac{M_{\mathrm{em}}}{M_{\mathrm{enc}}} \right )^3 \right ]
	\label{eq:tau_a_sca}
\end{equation}

We want to know at which embryo mass $ M_{\mathrm{em}} $ will Type -- I dominate the planetesimal -- driven migration.  Equating Eq.~\ref{eq:tau_a_type_I_e_zero} and Eq.~\ref{eq:tau_a_sca}, and solving the resulting cubic equation for $ M_{\mathrm{em}} $, we find that this occurs when $ M_{\mathrm{em}} \geq M_{\mathrm{em,crit}} $:

\begin{equation}
	M_{\mathrm{em,crit}} \equiv A F(a) - B \dfrac{G(a)}{F(a)}
	\label{eq:mem_crit}
\end{equation}
\noindent where the functions $ F(a) $ and $ G(a) $ are given below:

\begin{subequations}
	\label{eq:mem_crit_fns}
	\begin{align}
		F(a) &= \left [ C_1 H(a) + C_2\sqrt{C_3 H^2(a) + C_4 G^3(a)} \right ]^{1/3} \label{eq:mem_crit_f} \\
		G(a) &= \left ( \frac{f_s}{1.0} \right )^3 \left ( \frac{a}{\hbox{5.0 AU}} \right )^{3(2 - \alpha)} \label{eq:mem_crit_g} \\
		H(a) &= \left ( \frac{f_s}{1.0} \right )^3 \left ( \frac{c_a}{1.0} \right )^{-1} \left ( \frac{f_{gs}}{60.0} \right )^{-1} \left ( \frac{h_0}{\hbox{0.05}} \right )^2 \left ( \frac{a}{\hbox{5.0 AU}} \right )^{(13 - 6\alpha)/2} \label{eq:mem_crit_h}
	\end{align}
\end{subequations}
\noindent where $ f_{gs} $ is the gas to solid ratio, $ h_0 $ is the aspect ratio of the disk at 1.0 AU, and the coefficients are listed in Table~\ref{tab:model_coeff}.\\

{
\renewcommand{\baselinestretch}{1}
\small\normalsize

\begin{table}[ht!]
	\begin{center}
	\begin{tabular}{c c c c c c}
	\hline\hline
	$ A \, (\MEarth) $ & $ B \, (\MEarth) $ & $ C_1 $ & $ C_2 $ & $ C_3 $ & $ C_4 $ \\
	\hline
	$ 0.3816 $  & $ 0.2522 $  & $ 26.50 $ & $ 1.732 $ & $ 2.341 \times 10^{2} $ & $ 9.623 \times 10^{-2} $ \\
	\hline
	\end{tabular}
	\caption[model_coeff]{\label{tab:model_coeff}
	\label{lasttable}			
		Coefficients for Eq.~\ref{eq:mem_crit} and Eq.~\ref{eq:mem_crit_fns}.}
	\end{center}
\end{table}
}

\begin{figure}[ht!]
	\begin{center}
	\includegraphics[width = \columnwidth]{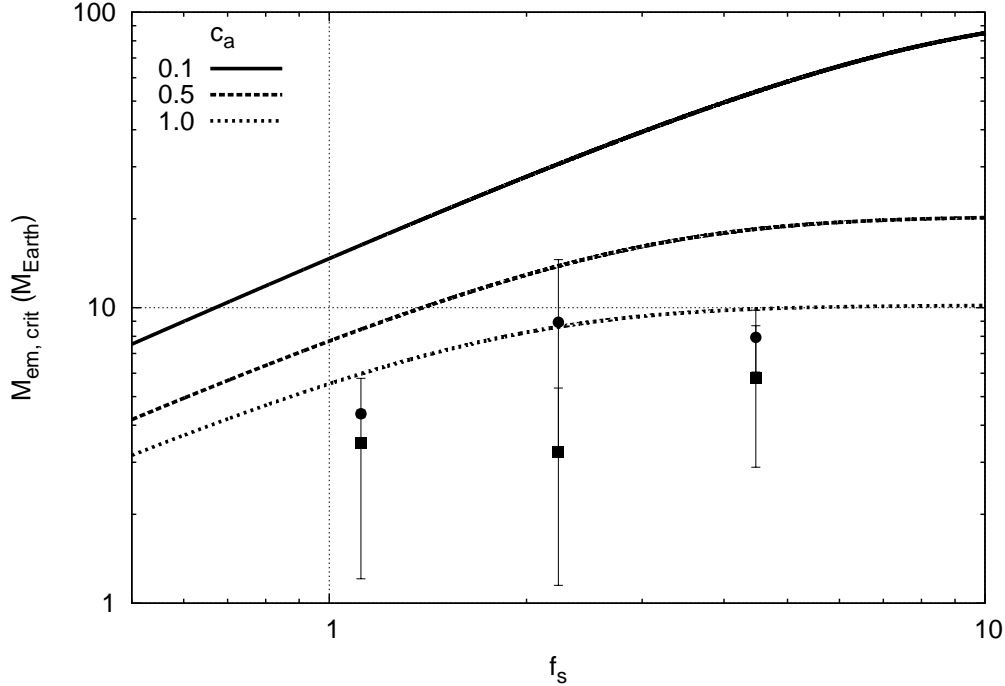}
	\caption[mem_crit_fs]{\label{fig:mem_crit_fs}
			\label{lastfig}			
		The critical embryo mass as a function of the solid surface density scaling factor $ f_s $, above which the Type -- I migration timescale is shorter than the planetesimal -- driven migration timescale (cf. Eq.~\ref{eq:mem_crit}).  This assumes $ f_g = f_s $ for a disk model with $ \alpha = 9/4 $ at 5.0 AU.  The curves correspond to $ M_{\mathrm{em,crit}}$ for three different values of of the Type -- I efficiency factor $ c_a $, while the points are the average maximum mass attained from simulations.  The simulations start with a $ 0.25 \, \MEarth $ embryo at 5.0 AU, and where the circles and squares are for simulations with $ c_a = 0.1 $ and $ c_a = 0.5 $, respectively.}
	\end{center}
\end{figure}

To test the predicted values of $ M_{\mathrm{em,crit}} $, we performed 8 sets of simulations of an embryo migrating outwards for several different $ f_s $ and $ c_a $ values, accreting planetesimals as it migrates.  We assess the critical mass to be when the embryo ceases to migrate outward, and reverses direction.  The simulations start with a $ 0.25 \, \MEarth $ embryo at 5.0 AU, with a population of $ 6.4 \times 10^4 $ dynamically cold (\ie $ \sigma_e = 2\sigma_I = 10^{-3} $) planetesimals distributed from 5.0 AU to 15.0 AU.  All these simulations are integrated for $ 10^5 \, \hbox{year} $ with a 0.5 year time step, and we assume the Type -- I efficiency parameter for eccentricity is set to $ c_e = 1.0 $.  Plotted in Fig.~\ref{fig:mem_crit_fs} is the maximum embryo mass attained for disk models with $ f_s = \{0.5, 1.0, 2.0 \}\sqrt{5} $ and $ \alpha = 9/4 $, each for $ c_a = \{ 0.1, 0.5 \} $.

First we note that the typical maximum mass attained is below the desired $ 10.0 \, \MEarth $, however the range of maximum mass does approach this limit for larger $ f_s $ values, and with marginal dependence on $ c_a $.  We also note that the maximum mass attained in these simulations do not exceed the predicted values from Eq.~\ref{eq:mem_crit}, even within the error bars.  This may be a consequence of the relatively low efficiency of planetesimal accretion, or that the criterion does not capture all the dynamics.  However in LTD10 they demonstrated that the addition of an atmosphere can dramatically increases the accretion efficiency, which is a promising prospect.  Though this atmosphere code can be made available, we shall leave the inclusion of atmospheres for future work.

We have omitted the $ c_a = 1.0 $ case in this investigation since it can be demonstrated that for a viscously evolving disk, one would need to consider an embryo mass an order of magnitude smaller than investigated in this study.  This can be demonstrated by using the results found in \citet{2010ApJ...715L..68L}, where simulations of viscously and radiatively evolving protoplanetary disks are presented.  From these simulations (See their Fig. 3) we assess the viscous accretion timescale of their disks to be on the order of $ \sim 5 \, \hbox{Myr} $, and if this timescale is compared to the Type -- I timescale in Eq.~\ref{eq:tau_a_type_I_e_zero}, we can place a constraint on $ M_{\mathrm{em}} $ or $ c_a $.  Assuming the fiducial values for the disk model we consider in our simulations, this gives $ M_{\mathrm{em}} \simeq 0.02 \, \MEarth/c_a $.  This relation implies that we would need to consider a $ 0.02 \, \MEarth $ embryo if $ c_a = 1.0 $, which is an embryo mass an order of magnitude smaller than we consider in this work.

As a final note, while the eventual dominance of Type -- I migration does not solve the problem of cores being deposited onto the central star, it is important to note the location of the giant planetary core when Type -- I migration begins to dominate.  Since planetesimal -- driven migration can migrate an embryo several AU further away from the central star, this can significantly increase the timescale for the giant planetary core to reach the central star.  For the example illustrated in Fig.~\ref{fig:outward_migration_growth}, tripling the embryo's initial distance to $ \sim 14.5 \, \hbox{AU} $ increases the Type -- I migration timescale in Eq.~\ref{eq:tau_a_type_I_e_zero} by more than an order of magnitude (\eg $ \tau_{\mathrm{a,typeI}} \sim 1.4 \, \hbox{Myr} $).  This may be sufficient to permit the giant planetary core to survive long enough for it to accrete a massive gaseous envelope, and with the dispersal of the gaseous disk decrease the effectiveness of Type -- I migration to deposit planets onto the central star.

\section{Conclusions\label{conclusions}}

To explain the observed distribution of giant extrasolar planets, including the giant planets in the Solar System, a comprehensive model of their formation is needed.  However, the physical processes and conditions involved give rise to very complex dynamical behaviour.  In view of the results of LTD10, a detailed study of planetesimal scattering and gas drag fills a much needed niche.  The resulting dynamics for a single embryo embedded in a sea of planetesimals subject to aerodynamic drag is quite fascinating and provides clues to future avenues of research.

The findings of this research are:
\begin{enumerate}
\item The presence of a gas disk modifies planetesimal -- driven migration of a single embryo, resulting in four different migration regimes: streaming, gap clearing, outward and inward embryo migration.
\item The range of planetesimal radii that result in outward embryo migration are plausible, and represents a possible dynamical avenue for embryo growth.
\item There exists simple scaling relations for the transition between the streaming and gap clearing regimes, along with the outward and inward embryo migration regimes.
\item Our toy model provides a good qualitative understanding of the observed phenomena in the $ N $-body simulations.
\item Simulations with a static planetesimal size distribution also demonstrate the different migration outcomes, and it was shown that outward migration is possible for a finite range of the power-law exponent of the size distribution.
\item Inclusion of Type -- I migration with planetesimal -- driven migration and aerodynamic gas drag were explored.  An analytical estimate was shown of the maximum achievable embryo mass before Type -- I migration begins to dominate over outward planetesimal -- driven migration, which appears to be borne out in simulations. 
\end{enumerate}

As described in a companion paper (LTD10), future work will incorporate other effects: multiple embryos, evolving planetesimal size distribution and embryo atmospheres, and the effects of fragmentation.


\ack
The ongoing financial support of NSERC is gratefully acknowledged.  HFL is grateful for funding from NASA's Origins and OPR programs, as well as a grant from the National Science Foundation (Award ID 0708775).

\label{lastpage}


\bibliography{capobianco10}

\bibliographystyle{elsart-harv}

\end{document}